\documentclass[aps, prl, floatfix, superscriptaddress,twocolumn,reprint]{revtex4-2}

\usepackage{amsmath}
\usepackage{amsfonts}
\usepackage{graphicx}
\usepackage{bm}
\usepackage{dsfont}
\usepackage{xfrac}
\usepackage{siunitx}
\usepackage{nicefrac} 
\usepackage{hyperref}
\usepackage[normalem]{ulem}
\usepackage{braket}
\usepackage{float}

\usepackage[dvipsnames]{xcolor}

\hypersetup{
    colorlinks=true,
    linkcolor=OliveGreen,
    urlcolor=BrickRed,
    citecolor=Blue,
    breaklinks=true,
    pdftitle={Probing Majorana wavefunctions in Kitaev honeycomb spin liquids with second-order two-dimensional spectroscopy},
    pdfauthor={Yihua Qiang, Victor L. Quito, Tha\'{i}s V. Trevisan, Peter P. Orth}
}

\definecolor{forestgreen}{rgb}{0.13, 0.55, 0.13}

\newcommand{\bfu}{\mathbf{u}}

\begin{document}
\title{Probing Majorana wavefunctions in Kitaev honeycomb spin liquids with second-order two-dimensional spectroscopy}

\author{Yihua Qiang}
\affiliation{Department of Physics and Astronomy, Iowa State University, Ames,
Iowa 50011, USA}
\affiliation{Ames National Laboratory, Ames, Iowa 50011, USA}

\author{Victor L. Quito}
\affiliation{Department of Physics and Astronomy, Iowa State University, Ames,
Iowa 50011, USA}
\affiliation{Ames National Laboratory, Ames, Iowa 50011, USA}

\author{Tha\'{i}s V. Trevisan}
\altaffiliation{Present address: Materials Sciences Division, Lawrence Berkeley National Laboratory, Berkeley, California 94720, USA}
\affiliation{Department of Physics and Astronomy, Iowa State University, Ames,
Iowa 50011, USA}
\affiliation{Ames National Laboratory, Ames, Iowa 50011, USA}

\author{Peter P. Orth} 
\affiliation{Department of Physics and Astronomy, Iowa State University, Ames,
Iowa 50011, USA}
\affiliation{Ames National Laboratory, Ames, Iowa 50011, USA}
\affiliation{Department of Physics, Saarland University, 66123 Saarbr\"ucken, Germany}

\date{\today}

\begin{abstract}
Two-dimensional coherent terahertz spectroscopy (2DCS) emerges as a valuable tool to probe the nature, couplings, and lifetimes of excitations in quantum materials. It thus promises to identify unique signatures of spin liquid states in quantum magnets by directly probing properties of their exotic fractionalized excitations. Here, we calculate the second-order 2DCS of the Kitaev honeycomb model and demonstrate that distinct spin liquid fingerprints appear already in this lowest-order nonlinear response $\chi^{(2)}_{yzx}(\omega_1, \omega_2)$ when using crossed light polarizations. We further relate the off-diagonal 2DCS peaks to the localized nature of the matter Majorana excitations trapped by $\mathbb{Z}_2$ flux excitations and show that 2DCS thus directly probes the inverse participation ratio of Majorana wavefunctions. By providing experimentally observable features of spin liquid states in the 2D spectrum, our work can guide future 2DCS experiments on Kitaev magnets. 
\end{abstract}

\maketitle

\emph{Introduction.--} Spectroscopic techniques are among the most powerful interrogation methods of quantum materials by directly measuring electronic Green's functions~\cite{dresselElectrodynamicsSolidsOptical2002, basovElectrodynamicsCorrelatedElectron2011, devereauxInelasticLightScattering2007,sobotaAngleresolvedPhotoemissionStudies2021,mukamelPrinciplesNonlinearOptical1999}. While much insight can be gained in linear response, nonlinear response functions often provide a wealth of additional information that is inaccessible in the linear regime. Examples include nonlinear conductivities that probe the Berry phase and quantum geometry of the electronic wavefunction in solids~\cite{sodemannQuantumNonlinearHall2015, maObservationNonlinearHall2019, laiThirdorderNonlinearHall2021,ahnRiemannianGeometryResonant2022} and second-harmonic generation that is extremely sensitive to a system's symmetry~\cite{fiebigSecondharmonicGenerationTool2005,zhaoSecondHarmonicGeneration2018,siricaPhotocurrentdrivenTransientSymmetry2022}. Another striking example is two-dimensional coherent spectroscopy (2DCS), which exposes the system to a sequence of coherent light pulses in order to measure a higher-order retarded Green's function~\cite{mukamelPrinciplesNonlinearOptical1999,hammConceptsMethods2D2011,luTwoDimensionalSpectroscopyTerahertz2018}. It provides a detailed two-dimensional excitation map of two frequencies that can be used to extract the nature, couplings and lifetimes of elementary excitations. This technique has long been used in the radio and optical frequency range and has only recently been extended to terahertz (THz) frequencies, which are ideal for the study of excitations and collective modes in quantum materials~\cite{kuehnTwoDimensionalTerahertzCorrelation2011, woernerUltrafastTwodimensionalTerahertz2013,bowlanUltrafastTerahertzResponse2014,luCoherentTwoDimensionalTerahertz2017,johnsonDistinguishingNonlinearTerahertz2019,mahmoodObservationMarginalFermi2021,linMappingMathrmLiNbO32022,luoQuantumCoherenceTomography2022}.  
\begin{figure}[b]
    \centering
\includegraphics[width=\linewidth]{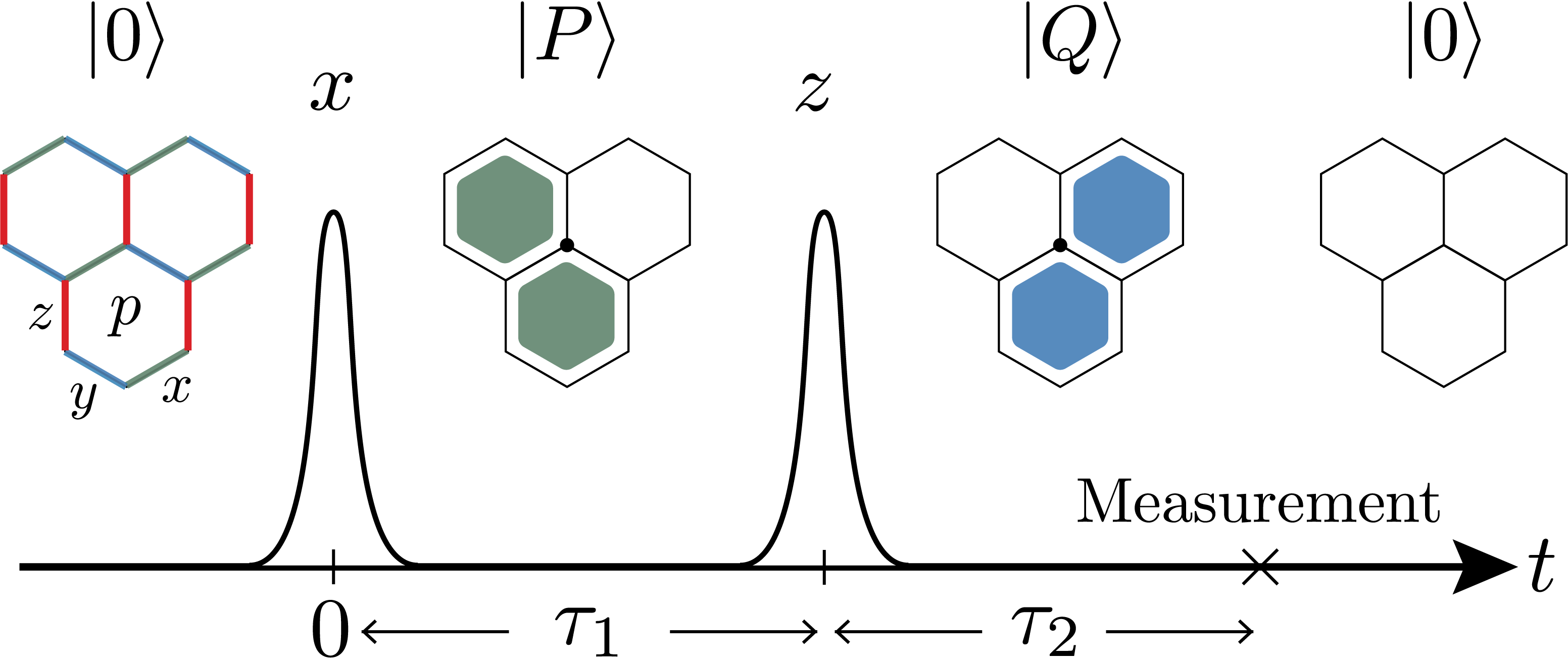}
    \caption{Magnetic field pulse sequence $B(t)$ used to measure $\chi^{yzx}(\tau_1, \tau_2)$ in the Kitaev honeycomb model. Different bond colors denote the $a$-bonds ($a=x,y,z$) and $p$ labels plaquettes. 
    The figure shows effect of the pulses on the flux configuration in the $R_1$ process. 
    Initially, the system is in the flux-free ground state $\ket{0}$, when at $t=0$ an $x$-polarized pulse creates a pair of $x$-fluxes (green) next to a spin at site $j$ (black dot), resulting in state $\ket{P}$. At $\tau_1$, a $z$-polarized pulse creates a pair of $z$-fluxes. Since the system needs to return to the flux-free state in the end, the $z$-bond must be connected to the same spin $j$, resulting in a $y$-flux pair (blue) in state $\ket{Q}$. Measurement of the magnetization $M^y(\tau_1 + \tau_2)$ removes the $y$-flux pair and system returns to a flux-free state.}
    \label{fig:1}
\end{figure}

Being able to disentangle different types of excitations and to discriminate between intrinsic and inhomogeneous broadening, THz 2DCS has been proposed to provide unique fingerprints of fractionalized excitations in exotic quantum magnets~\cite{wanResolvingContinuaFractional2019,nandkishoreSpectroscopicFingerprintsGapped2021,parameswaranAsymptoticallyExactTheory2020,choiTheoryTwoDimensionalNonlinear2020,liPhotonEchoLensing2021}. A previous theoretical study of 2DCS in the Kitaev honeycomb spin liquid~\cite{choiTheoryTwoDimensionalNonlinear2020}, for example, has shown that the third-order diagonal susceptibility $\chi^{(3)}_{zzzz}$ contains signatures of the two types of fractionalized excitations in the Kitaev model: static $\mathbb{Z}_2$ gauge fluxes and itinerant Majorana fermion excitations. Here, we demonstrate that marks of fractionalization are already present in the lower second-order off-diagonal response tensor element $\chi^{(2)}_{yzx}$, which is much larger in intensity and thus experimentally easier accessible. We find clear evidence of the presence of a nonzero flux gap and a broad continuum of Majorana fermion excitations, whose intrinsic lifetimes can be extracted from the 2D spectrum.
In addition, we show that $\chi^{(2)}$ provides direct evidence of the trapping of Majorana wavefunctions around static $\mathbb{Z}_2$ flux excitations and that the ratio of second and first-order response, $\chi^{(2)}/\chi^{(1)}$, is a quantitative measure of the overlap of such localized Majorana wavefunctions. Our work thus directly links localized Majorana states trapped around $\mathbb{Z}_2$ gauge fluxes to observable peaks in the 2D spectrum, and we relate the inverse participation ratios of the wavefunctions to the peak sizes. Finally, we show how exchange anisotropies modify the 2D spectrum, which can be used as a sensitive experimental probe of anisotropies. 

Identifying unique fingerprints of spin liquid states with 2DCS promises to become a fruitful direction in the experimental study of Kitaev magnets~\cite{savaryQuantumSpinLiquids2017, trebstKitaevMaterials2022}. Anisotropic compass-like Kitaev spin interactions are found in $d$-electron materials with strong crystal field and spin-orbit interactions~\cite{chaloupkaKitaevHeisenbergModelHoneycomb2010, kimchiKitaevHeisenbergModelsIridates2014,liuPseudospinExchangeInteractions2018, sanoKitaevHeisenbergHamiltonianHighspin2018}. 
Proposals for possible realizations of a Kitaev spin liquid on the honeycomb lattice include $\alpha\text{-RuCl}_{3}$~\cite{banerjeeProximateKitaevQuantum2016, doMajoranaFermionsKitaev2017,suzukiProximateFerromagneticState2021}, iridates~\cite{singhRelevanceHeisenbergKitaevModel2012, williamsIncommensurateCounterrotatingMagnetic2016, revelliFingerprintsKitaevPhysics2020} and cobaltates~\cite{liuKitaevSpinLiquid2020, zhangMagneticContinuumCobaltbased2023, halloranGeometricalFrustrationKitaev2023, tuEvidenceGaplessQuantum2023}.
The main challenge is to differentiate the phenomena associated with the Kitaev exchange from those due to Heisenberg and other exchange interactions. While the latter often drive the system into a magnetically ordered ground state, unusual spin-liquid-like behavior has been observed in the presence of a magnetic field. To this end, we here calculate the second-order 2DCS response of the pure Kitaev honeycomb model in order to provide clear signatures of the spin liquid state that can guide experimental studies of Kitaev magnets.  

\emph{Kitaev model.--}
The ferromagnetic Kitaev spin model on the honeycomb lattice is defined as~\cite{kitaevAnyonsExactlySolved2006},
\begin{equation}
\label{eq:ham_1}
H =-J_{x} \sum_{\langle i,j\rangle_x} \sigma_{i}^{x} \sigma_{j}^{x}-J_{y} \sum_{\langle i,j\rangle_y} \sigma_{i}^{y} \sigma_{j}^{y}-J_{z} \sum_{\langle i,j\rangle_z} \sigma_{i}^{z} \sigma_{j}^{z}\,.
\end{equation}
Here, $J_a > 0$ and $\sigma_j^{{a}}$ with $a=x,y,z$ represent Pauli matrices at site $j$ of the honeycomb lattice, which has two basis sites $A$ and $B$ per unit cell. Each spin has three nearest-neighbors, and $\langle i,j \rangle_a$ sums over nearest-neighbor pairs connected by an $a$-bond (see Fig.~\ref{fig:1}). 
The Kitaev model is exactly solvable because every honeycomb plaquette $p$ hosts a flux operator $\hat{W}_p={\sigma}_1^x {\sigma}_2^y {\sigma}_3^z {\sigma}_4^x {\sigma}_5^y {\sigma}_6^z$ ($j=1,\ldots,6$ label the sites around the plaquette) that commutes with the Hamiltonian and with all other $W_{p'}$. The flux operator has eigenvalues $w_p = \pm 1$ and a plaquette is flux-free if $w_p = +1$ and has a flux otherwise. 

Kitaev's solution involves writing the spin operators $\sigma_j^{{a}} = i b_j^{{a}} c_j$ using four Majorana fermions $b_j^{x}, b_j^{y}, b_j^{z}, c_j$, which satisfy $\{b_{i}^{{a}}, b_{j}^{d}\}=2 \delta_{i j}\delta_{ad}$, $\left\{c_{i}, c_{j}\right\}=2 \delta_{i j}$, and $\{b_{i}^{{a}}, c_j\}=0$. One refers to $b_j^a$ as bond fermions and to $c_j$ as matter fermions. The introduction of four Majoranas per site doubles the Hilbert space and leads to a local $\mathbb{Z}_{2}$ gauge field, which poses the main challenge when computing correlation functions~\cite{baskaranExactResultsSpin2007, knolleDynamicsTwoDimensionalQuantum2014, choiTheoryTwoDimensionalNonlinear2020}. The constraint $D_j = b_j^x b_j^y b_j^z c_j = \mathds{1}\, \forall j$ restores the physical Hilbert space. In terms of Majorana fermions, the spin Hamiltonian takes the form $H_{\hat{u}} = \frac{i}{2} \sum_{j, k} \hat{A}_{j k} c_{j} c_{k}$. Here, $\hat{A}_{j k} = J_{{a}} \hat{u}_{j k}$ if $j$ and $k$ sites are connected by an ${a}$-bond and zero otherwise. The bond operators $\hat{u}_{j k} = i b_{j}^{{a}} b_{k}^{{a}}$ ($j$ is always an $A$ site) commute with the Hamiltonian and among themselves. They can thus be replaced by their eigenvalues $u_{jk} = \pm 1$ and a particular bond configuration $\bfu \equiv \{u_{ij}\}$ determines the gauge-independent fluxes via $\hat{W}_p=\prod_{\langle j,k\rangle \in \partial p} \hat{u}_{jk}$. 

Since the fluxes are static, we can work in a particular gauge configuration $\bfu$, where the Hamiltonian $H_{\bfu}$ is quadratic in matter fermions $c_j$. Even though the matter spectrum and eigenstates depend on $\bfu$, it is convenient to write the eigenstates using a tensor product notation as $\ket{\psi} = \ket{F} \otimes \ket{M^\bfu}$ with the flux state $\ket{F} \equiv \ket{\{W_p\}}$ set by $\bfu$ and the matter state $\ket{M^\bfu}$ consisting of matter excitations on top of the vacuum $\ket{M_0^{\bfu}}$.
It is useful to introduce complex bond fermions as $\chi^a_{jk} = \frac{1}{2}(b^a_j + i b^a_k)$ 
, where $j$ is an $A$ site and $k$ is connected to $j$ by an $a$-bond. We choose the convention $\ensuremath{\hat{u}_{jk}=2(\chi^{a}_{jk})^{\dagger} \chi^{a}_{jk}-1}$ where the flux free state $\ket{F_0}$ corresponds to all bond fermions occupied and a general flux state reads $\ket{F} =  \chi_{j_n k_n}^{a_n}\cdots \chi_{j_1 k_1}^{a_1} \ket{F_0}$. Once the gauge field state is determined, one can diagonalize the matter part $H_\bfu = \sum_\lambda \epsilon_\lambda [2 (a^\bfu_\lambda)^\dag a^\bfu_\lambda -1 ]$ in terms of complex fermion eigenmodes $a^\bfu_\lambda$ and write its state as $\ket{M^\bfu} =(a^\bfu_{\lambda_s})^{\dag}\cdots(a^\bfu_{\lambda_1})^{\dag} \ket{M^\bfu_0}$ with vacuum $\ket{M^\bfu_0}$ (see Supplementary Material~\cite{qiangSupplementaryMaterial2023} for details).

\begin{figure}[t]
    \centering
    \includegraphics[width=0.95\linewidth]{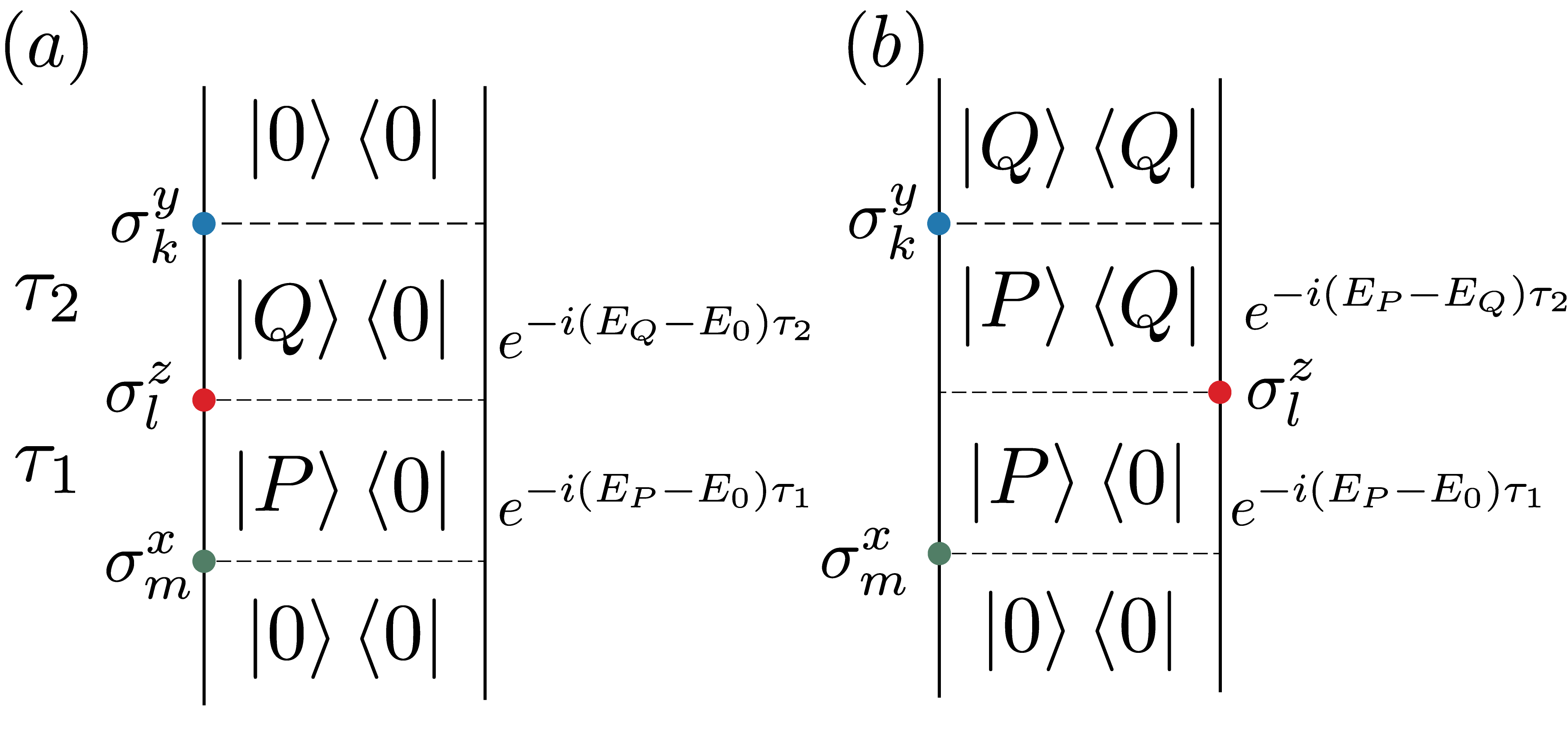}
    \caption{Liouville pathways for (a) $R_{1}$ and (b) $R_{2}$ processes. Time evolves from bottom to top, and dots represent bra or ket operations on the density matrix by the Pauli operators (summation over sites $m,l,k$ is done in the end). $\ket{0}$ is the ground state; $\ket{P}$ and $\ket{Q}$ denote excited states of the Hamiltonian, and the exponentials describe the phases acquired during time evolution over intervals $\tau_1$ and $\tau_2$. 
    }
    \label{fig:2}
\end{figure}

\emph{Second-order 2DCS response.--}
The ground state of Eq.~\eqref{eq:ham_1} is a spin liquid, which is gapless for $J_x + J_y > J_z$ (and permutations) and gapped otherwise. In the following we focus on the isotropic point $J_{x} = J_{y} = J_{z} \equiv J$ (see~\cite{qiangSupplementaryMaterial2023} for anisotropic couplings) and compute the second-order 2DCS response
\begin{equation}
\label{eq:chi2}
\begin{split}
    &\chi^{abc}(\tau_1,\tau_2)\\
    &~~=\frac{i^2}{{N}}\theta(\tau_1)\theta(\tau_2)\langle [[M^a(\tau_1+\tau_2),M^b(\tau_1)],M^{c}(0)]\rangle \,.
\end{split}
\end{equation}
Here, $N$ is the total number of unit cells and $M^{a}(t)=\sum_{j} \sigma_j^{a} (t)$ is the $a$-th component of the total magnetization in the Heisenberg picture, and the Heaviside $\theta$ functions guarantee the causality of the response. The expectation value is taken in the many-body ground state. This response corresponds to the nonlinear part of the magnetization $M^a$ induced by a sequence of two magnetic field pulses $B(t) = B_1^{c}\delta(t) + B_2^{b}\delta(t-\tau_1)$ shown in Fig.~\ref{fig:1}~\cite{woernerUltrafastTwodimensionalTerahertz2013,wanResolvingContinuaFractional2019}.
The second-order susceptibility $\chi^{abc}$ is finite only when $a,b,c$ are all different and  we consider $\chi^{yzx}$ in the following. At the isotropic point, all other nonzero components can be related by symmetry~\cite{qiangSupplementaryMaterial2023}. Expansion of the commutator in Eq.~\eqref{eq:chi2} shows that $\chi^{abc}$ consists of two contributions,
\begin{equation*}
    \chi^{yzx}(\tau_1,\tau_2)=-\frac{1}{{N}}\theta(\tau_1)\theta(\tau_2)[R_{1}(\tau_1,\tau_2) - R_{2}(\tau_1,\tau_2) + c.c.],
\end{equation*}
where
\begin{equation}
\begin{split}
R_{1}(\tau_1,\tau_2) &= \langle M^{y}(\tau_1+\tau_2)M^{z}(\tau_1)M^{x}(0)\rangle, \\
R_{2}(\tau_1,\tau_2) &= \langle M^{z}(\tau_1) M^{y}(\tau_1+\tau_2)M^{x}(0)\rangle. \\
\end{split}
\end{equation}
We represent the processes in $R_{1}$ and $R_{2}$ with the Liouville pathways shown in Fig.~\ref{fig:2}~\cite{mukamelPrinciplesNonlinearOptical1999}. 
The system starts in the ground state density matrix $\ket{0}\bra{0}$ with energy $E_0$, where $\ket{0}$ is constructed with zero flux and matter fermions. We note that while this is not a physical state for our choice of periodic boundary conditions and geometry, which is required to contain one matter fermion $\left|0\right\rangle _{\text{phys}} = \ket{F_0}\otimes a^{\dag}_{1} \ket{M_0}$~\cite{zschockePhysicalStatesFinitesize2015}, it is well known that physical and unphysical states yield identical results for large enough system size~\cite{zschockePhysicalStatesFinitesize2015, choiTheoryTwoDimensionalNonlinear2020}. Using the zero matter ground state reduces the complexity of the calculations and facilitates the interpretation of the results.
Since the entire spectrum of the Kitaev Hamiltonian~\eqref{eq:ham_1} is known, we can use the Lehmann representation and insert two resolutions of identity $\sum_{P} \ket{P}\bra{P} =\sum_{Q} \ket{Q}\bra{Q} = 1$: 
\begin{align}
\chi^{yzx}_{R_1}&=\frac{-2}{{N}} \text{Re} \sum_{PQ} \sum_{klm}  \braket{0|\sigma_k^y| Q} \braket{Q|\sigma_l^z|P}\braket{P|\sigma_m^{x}| 0} \nonumber \\ 
    &  \times\theta(\tau_1)\theta(\tau_2) e^{-i\tau_1(E_P-E_0)} e^{-i\tau_2(E_Q-E_0)} \,,
    \label{eqn:R1}
\end{align}
\begin{align}
\chi^{yzx}_{R_2}&=\frac{2}{{N}}\text{Re} \sum_{PQ} \sum_{klm}  \braket{0| \sigma_l^z|Q} \braket{Q |\sigma_k^y| P}\braket{P| \sigma_m^{x}| 0} \nonumber \\ 
    & \times\theta(\tau_1)\theta(\tau_2) e^{-i\tau_1(E_P-E_0)} e^{-i\tau_2(E_P-E_Q)}\,.
    \label{eqn:R2}
\end{align}
Here, each pathway is combined with its time-reversed partner as $\chi^{yzx}_{R_n}(\tau_1,\tau_2) = \theta(\tau_1)\theta(\tau_2)[R_{n}(\tau_1,\tau_2)+R_{n}^{*}(\tau_1,\tau_2)]/{N}$, and the states $\ket{P}$ and $\ket{Q}$ are eigenstates of the Hamiltonian~\eqref{eq:ham_1} with energy $E_P$ and $E_Q$.

The states $\ket{P}$ are connected to the flux-free ground state $\ket{0}$ and the first pulse at $t=0$ is polarized in the $x$-direction. Therefore, nonzero matrix elements only occur if $\ket{P}$ contains a pair of $x$-fluxes $\ket{P} = \chi_{mn}^{x}\ket{F_0} \otimes (a^\bfu_{\lambda_s})^{\dag} \cdots (a^\bfu_{\lambda_1})^{\dag} \ket{M_0^\bfu}$. Here, $\ket{M_0^\bfu}$ is the vacuum for $\bfu$ with one $x$ bond flipped, $u_{mn} = -1$. A phase $e^{-i\tau_1(E_P-E_0)}$ is acquired during the time evolution by $\tau_1$. Note that we truncate the matter fermion number in the intermediate state $\ket{P}$ to one, which is known to be an excellent approximation~\cite{choiTheoryTwoDimensionalNonlinear2020}. 
Next, a pulse polarized in the $z$-direction arrives at time $\tau_1$ and creates a pair of $z$-fluxes at site $l$ via application of $\sigma^z_l$. Since after the measurement of $\sigma^y_k$ at time $\tau_1 + \tau_2$ the system must return to either the initial state (for pathway $R_1$) or a diagonal state $\ket{Q}\bra{Q}$ (for $R_2$), the sites $l, k$  must be in proximity to site $m$ such that the fluxes overlap and partially annihilate each other~\cite{qiangSupplementaryMaterial2023}. As a result, the state $\ket{Q}$ contains a pair of $y$-fluxes when computing $R_1$ and a pair of $z$-fluxes that can be obtained by application of $\sigma^y_k$ to $\ket{P}$ for pathway $R_2$. We also truncate the number of matter fermions in state $\ket{Q}$ to be maximally one. 
It is worth highlighting another difference between the $R_1$ and $R_2$ processes. For $R_1$ the $z$-polarized pulse induces a ket operation, leading to a transition from $\ket{P}$ to state $\ket{Q}$ and the phase acquired during $\tau_2$ is $e^{-i \tau_2(E_Q - E_0)}$. In contrast, for $R_2$ the $z$-polarized pulse induces a bra operation onto the density matrix and creates a coherence $\ket{P}\bra{Q}$. The phase accumulated during time evolution $\tau_{2}$ is thus $e^{-i\tau_2(E_P-E_Q)}$.

\begin{figure}[t!]
  \includegraphics[width=\linewidth]{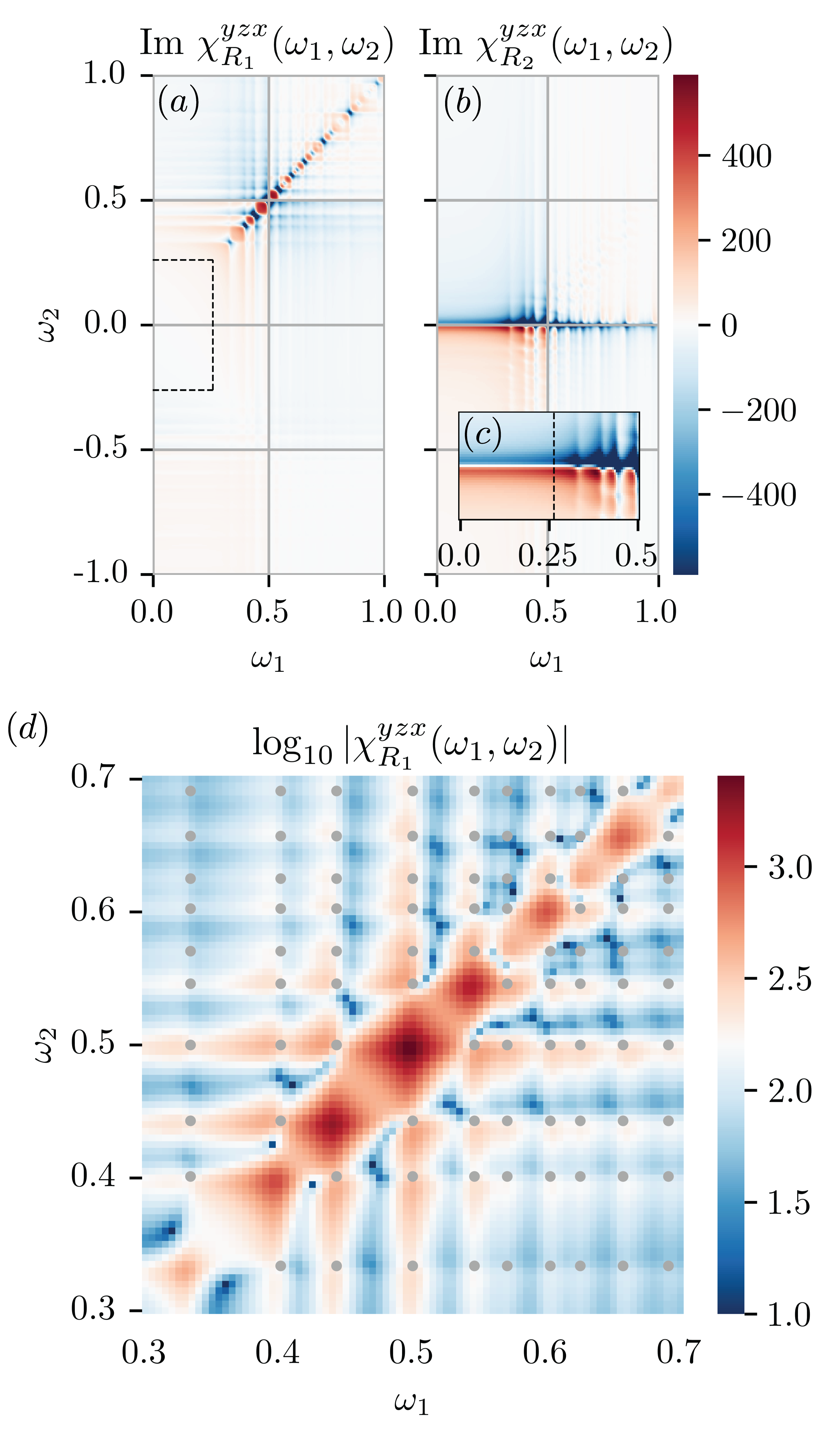}
  \caption{2D spectrum of second-order response $\chi^{yzx}(\omega_1,\omega_2)$ for $100 \times 100$ lattice, $N=10^4$, $J_a=1,\Gamma=0.01$. (a) Imaginary part of $\chi^{yzx}_{R_1}$ shows peaks along the diagonal $\omega_1 = \omega_2$. Vertical and horizontal streaks are due to the principal value parts from $g$ functions. The dashed box indicates the flux gap $E_g$ below which the response vanishes. (b) Imaginary part of $\chi^{yzx}_{R_2}$ exhibits vertical streaks at energies $\omega_1 = E_P - E_0$. (c) Inset zooms into low-energy region of panel (b). Below the flux gap (dashed line), no vertical streaks appear. (d) Absolute value of $\chi^{yzx}_{R_1}$ on a logarithmic scale in region $0.25 J \leq \omega_{1,2}\leq J$. Grey dots denote excitation energies of localized states with high IPR.}
  \label{fig:3}
\end{figure}

\emph{Results and Discussion.--} We analyze $\chi^{yzx}$ in frequency space and label by $\omega_1$ and $\omega_2$ the frequencies conjugate to the time intervals $\tau_1$ and $\tau_2$, respectively. 
The responses are written in terms of a product of matrix elements and the function $g(x) = i/(x+i\Gamma)$, with the broadening $\Gamma$ coming from the scattering of quasiparticles. The second-order response involves the product of two $g$-functions $g(x_{1})g(x_{2})$~\cite{qiangSupplementaryMaterial2023}. In the small $\Gamma$ limit, it leads to the terms $\delta(x_{1})\delta(x_{2})-\mathcal{P}\frac{1}{x_{1}}\mathcal{P}\frac{1}{x_{2}}$ in the imaginary part of the response, because the product of matrix elements is purely imaginary. The real part of the response contains terms mixing principle values and delta-like contributions $\delta(x_{1})\mathcal{P}\frac{1}{x_{2}}+\delta(x_{2})\mathcal{P}\frac{1}{x_{1}}$. 
Such mixing 
is a general feature in nonlinear response functions~\cite{nandkishoreSpectroscopicFingerprintsGapped2021, choiTheoryTwoDimensionalNonlinear2020}. Taking into account the time-reversal partners, we notice that the real parts of the 2D spectra are symmetric about the origin, while the imaginary part is antisymmetric.

\begin{figure}[t]
    \centering
    \includegraphics[width=.9\linewidth]{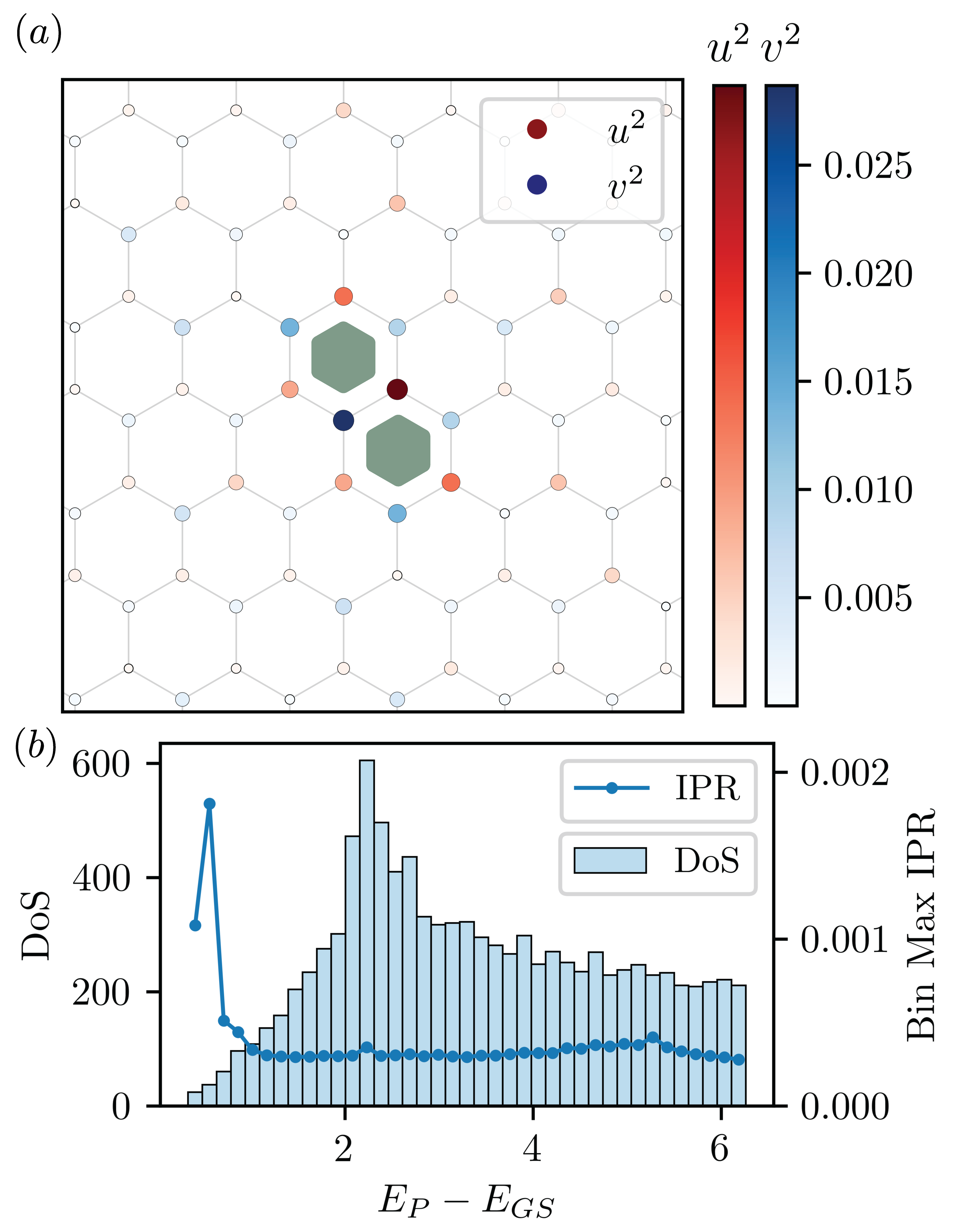}
    \caption{(a) A representative localized state $\ket{P}$ state trapped near an $x$-flux pair (green). The color and size of each site denote the amplitudes $u$ and $v$ of the Majorana matter wavefunction.  
    (b) Density of states and maximal IPR for states within a given energy bin. A few high IPR states (order $\sim 1/{N}$ of the total states) at low energies make up for most of the linear and nonlinear response.}
    \label{fig:4}
\end{figure}

We plot the 2D spectrum of $\text{Im} \chi^{yzx}(\omega_1, \omega_2)$ in Fig.~\ref{fig:3} for a lattice with $100 \times 100$ unit cells, $N = L^2 = 10^4$. Panel $(a)$ shows the contribution of pathway $R_1$ and panel (b) the one from pathway $R_2$ (see~\cite{qiangSupplementaryMaterial2023} for the real parts and the sum of both pathways). Given the symmetry properties of the response functions, we only show results for $\omega_1>0$. 
Panel $(c)$ shows $\chi^{yzx}_{R_2}$ in the frequency window $0.3 J<\omega_{1}, \omega_2<0.7 J$ on a logarithmic scale. 
We start by analyzing the results for pathway $R_{1}$. The peaks of the $R_1$ process appear near the diagonal $\omega_1=\omega_2$. Investigating the spectrum over a wider frequency range~\cite{qiangSupplementaryMaterial2023} shows that the largest response occurs in the shown frequency range. The dashed box indicates the flux gap $E_g= \text{min}( E_{P(Q)}) - E_{0} = 0.263 J$ in the thermodynamic limit, which is the minimal energy cost of excitations. Due to the product of $g$-functions, the peaks occur at $\omega_{1}= E_P-E_0$ and $\omega_{2}=E_Q-E_0$. As discussed above, for a given site $m$ in $M^{x}$ (or $M^y$), the states $P$ (or $Q$) have fluxes at honeycomb plaquettes neighboring site $m$ connected by $x$ (or $y$) bonds. Since we consider the isotropic case, $x$ and $y$ fluxes cost the same energy and the signal vanishes inside the region $|\omega_{1,2}|<E_{g}$. 
Interestingly, we find the strongest signal along the diagonal, centered around energies $\omega_{1}=\omega_2 \approx 0.5 J$, even though the joint density of states in this region is not large [see Fig.~\ref{fig:4}(b)]. This implies that the response is due to the matrix elements being large for these processes. Below, we show that it indeed derives from localized matter Majorana states that are trapped around plaquettes with nonzero flux. This is in sharp contrast to results of the third-order response functions, where a strong diagonal peak arises from a constructive interference effect~\cite{choiTheoryTwoDimensionalNonlinear2020}.

In Fig.~\ref{fig:3}(c), we plot the absolute value of $\chi_{R_1}^{yzx}$ on a logarithmic scale to highlight the presence of off-diagonal peaks, which are due to transitions between $\ket{P}$ and $\ket{Q}$ states with different energies. These peaks are only about a factor of ten smaller than the diagonal ones, which indicates the locality of those states. Otherwise the matrix element of a \emph{local} operator $\braket{P|\sigma_l^z|Q}$ could not be large between the orthogonal states $\ket{P}$ and $\ket{Q}$. 

We now analyze $R_{2}$ shown in Fig~\ref{fig:3}(b), which exhibits vertical stripes that are centered at energies $\omega_1 = E_P - E_0$. The reason is that peaks along the $\omega_2$-axis occur at energy \emph{differences} $E_Q - E_P$ and thus densely overlap. As detailed in the inset panel~(c), the signal below the flux gap $E_g$ stems purely from the principal values as we observe no vertical tails for $\omega_1 < E_g$. The strongest vertical streaks occur at the same energies $\omega_1$ as in $R_1$ and arise from the large overlap of localized Majorana states as we show next. 

To quantitatively characterize the localization of the matter Majorana wavefunctions we present their inverse participation ratio (IPR) in Fig.~\ref{fig:4}(b). The IPR is of order 1 for localized states and of order $1/L^2$ for extended states~\cite{qiangSupplementaryMaterial2023}. 
The IPR distribution separates into two regions with a few states at low energy having a much higher IPR. We find that these states are indeed localized around non-zero fluxes. Fig.~\ref{fig:4}(a) shows a representative example at energy $E \approx 0.5 J$. 
The dominance of the low-energy peaks in $\text{Im}\chi^{yzx}$ for $\omega_{1,2} \lesssim J$ and the presence of the off-diagonal peaks in this region can thus be understood in terms of the large wavefunction overlap matrix elements $\braket{P|\sigma^z_k|Q}$ of high-IPR states. 
We note that this also accounts for most of the peak intensity in linear response $\chi^{aa}$~\cite{qiangSupplementaryMaterial2023}, which exclusively probes the diagonal elements $\braket{P|\sigma^a_k|P}$. The second-order response additionally contains information about the off-diagonal matrix elements $\braket{Q|\sigma_{z}|P}$ with $P \neq Q$. By taking the ratio of second- to first-order response, $\chi^{(2)}/\chi^{(1)}$, we can extract the size of this element from experiment. Being only one order of magnitude smaller in the low-energy region is a clear indication of the localized nature of the Majorana wavefunctios at these energies as discussed above~\cite{qiangSupplementaryMaterial2023}.  

Finally, we briefly comment on results away from the isotropic case. If $J_{x} \neq J_{y}$ the center of the peaks in $\text{Im} \chi^{yzx}_{R_1}$ shift away from the diagonal, reflecting the different energy costs of creating $x$ and $y$ fluxes. This can be used as a sensitive probe of exchange anisotropies. The sharp features originating from the localized matter fermions are still present~\cite{qiangSupplementaryMaterial2023}. 

\emph{Conclusions.--}
A primary challenge in the experimental search for spin liquids is to find their unique and observable signatures, and 
one promising path is to directly probe properties of their fractionalized excitations.  
The Kitaev spin liquids host two different types of fractionalized excitations, $\mathbb{Z}_2$ fluxes and matter Majorana fermions, which are not clearly separable in linear response, where a broad continuum of excitations occurs above the flux gap.
In contrast, we demonstrate that they can be disentangled in the second-order nonlinear susceptibility $\chi^{abc}$ with non-repeating indices. In addition, off-diagonal peaks in the 2D spectrum directly indicate the presence of localized Majorana matter excitations trapped by fluxes, and the 2DCS peak sizes are quantitatively related to the IPRs of their wavefunctions.  
Involving the lowest nonlinear response, our proposal of using crossed-polarization pulses to probe the off-diagonal second-order susceptibility $\chi^{yzx}$ is the experimentally most straightforward way of using 2DCS to probe fractionalized excitations in Kitaev spin liquids. 

\begin{acknowledgements}
We acknowledge valuable discussions with N. Peter Armitage, Yueqing Chang, Elio Koenig, Milan Kornja\v ca, Ana-Marija Nedi\' c, Natalia Perkins, Nicholas Sirica, Yuriy Sizyuk, and Yuan Wan. V.L.Q., T.V.T., and P.P.O. acknowledge support from the Research Corporation for Science Advancement via P.P.O.'s Cottrell Scholar Award. Y.Q. was supported by the U.S. Department of Energy, Office of Science, National Quantum Information Science Research Centers, Superconducting Quantum Materials and Systems Center (SQMS) under the contract No.~DE-AC02-07CH11359. The research was performed at the Ames National Laboratory, which is operated for the U.S. Department of Energy by Iowa State University under Contract
No.~DE-AC02-07CH11358.
\end{acknowledgements}


%

\clearpage
\begin{widetext}
\section*{Supplemental Material \label{sec:Supp_Material}}
This Supplemental Material includes symmetry analysis of the second-order response tensor and the constraints among different components (Section~\hyperref[sec:symm]{S1}), a derivation of the higher-order response functions (Section~\hyperref[sec:ft]{S2}), and a detailed computation of the matrix elements (Section~\hyperref[sec:mat_elem]{S3}). By comparing the peak positions and their intensities, we demonstrate the role played by the novel matrix element (Section~\hyperref[sec:off_diag]{S4}). We also present results for the anisotropic case (Section~\hyperref[sec:aniso]{S5}). The definition and analysis of the inverse participation ratio of our problem are explained in (Section~\hyperref[sec:ipr]{S6}). 

\section{Symmetry analysis of the second-order correlation functions \label{sec:symm}}

The crystal symmetries, combined with time reversal, lead to the following relations that hold, in general, for any choice of the
couplings $J_{x},J_{y}$ and $J_{z}$,
\begin{align*}
\chi^{x,y,z}\left(\omega_{1},\omega_{2}\right) & =-\chi^{y,x,z}\left(\omega_{2},\omega_{1}\right),\\
\chi^{x,z,y}\left(\omega_{1},\omega_{2}\right) & =-\chi^{z,x,y}\left(\omega_{2},\omega_{1}\right),\\
\chi^{y,x,z}\left(\omega_{1},\omega_{2}\right) & =-\chi^{x,y,z}\left(\omega_{2},\omega_{1}\right),\\
\chi^{yzx}\left(\omega_{1},\omega_{2}\right) & =-\chi^{z,y,x}\left(\omega_{2},\omega_{1}\right),\\
\chi^{z,x,y}\left(\omega_{1},\omega_{2}\right) & =-\chi^{x,z,y}\left(\omega_{2},\omega_{1}\right),\\
\chi^{z,y,x}\left(\omega_{1},\omega_{2}\right) & =-\chi^{yzx}\left(\omega_{2},\omega_{1}\right).
\end{align*}

At the isotropic point, the symmetry constraints impose that there is only one independent component of the tensor, and all other components can be related to that one in the following way:

\begin{align*}
\chi^{x,y,z}\left(\omega_{1},\omega_{2}\right) & =-\chi^{x,z,y}\left(\omega_{1},\omega_{2}\right),\\
 & =-\chi^{y,x,z}\left(\omega_{1},\omega_{2}\right),\\
 & =\chi^{yzx}\left(\omega_{1},\omega_{2}\right),\\
 & =\chi^{z,x,y}\left(\omega_{1},\omega_{2}\right),\\
 & =-\chi^{z,y,x}\left(\omega_{1},\omega_{2}\right).
\end{align*}

\section{Fourier transformation of the response functions \label{sec:ft}}
Here, we perform the necessary Fourier transformation from the time to the frequency domain. Having these expressions analytically in frequency avoids the need for Fourier transformation numerically. In general, the response function $\chi^{abc}(\tau_1,\tau_2)$ in Lehmann representation can be separated into contributions based on pathways. The contribution from pathway 1 is 
\begin{equation}
\begin{split}
\chi^{abc}_{R_1}(\tau_1,\tau_2) &= -\frac{1}{{N}}\theta(\tau_1)\theta(\tau_2)[R^{(1)}(\tau_1,\tau_2)+R^{(1)}(\tau_1,\tau_2)^{*}]\\
	&=-\frac{1}{{N}}\theta(\tau_1)\theta(\tau_2) 2\Re \sum_{PQ} \sum_{klm}  \bra{0} \sigma_k^a \ket{Q} \bra{Q}\sigma_l^b\ket{P}\bra{P} \sigma_m^{c} \ket{0} e^{-i\tau_1(E_P-E_0)} e^{-i\tau_2(E_Q-E_0)}.
\end{split}
\end{equation}
while the contribution from pathway 2 is
\begin{equation}
\begin{split}
\chi^{abc}_{R_2}(\tau_1,\tau_2) &= +\frac{1}{{N}}\theta(\tau_1)\theta(\tau_2)[R^{(2)}(\tau_1,\tau_2)+R^{(2)}(\tau_1,\tau_2)^{*}]\\
	&=+\frac{1}{{N}}\theta(\tau_1)\theta(\tau_2) 2\Re \sum_{PQ} \sum_{klm}  \bra{0} \sigma_l^b \ket{Q} \bra{Q} \sigma_k^a \ket{P}\bra{P} \sigma_m^{c} \ket{0} e^{-i\tau_1(E_P-E_0)} e^{-i\tau_2(E_P-E_Q)},
\end{split}
\end{equation}
where $\ket{P}$, $\ket{Q}$ are eigenstates of the Hamiltonian.

In frequency space, introducing $g(x) = \frac{i}{x+i \Gamma}$, where $\Gamma$ takes into account the level broadening, we rewrite
\begin{equation}
\begin{split}
\chi^{abc}(\omega_1,\omega_2) &= \int_{-\infty}^{\infty}d\tau_1 \int_{-\infty}^{\infty}d\tau_2~ e^{i\omega_1\tau_1}e^{i\omega_2\tau_2} \chi^{abc}(\tau_1,\tau_2)\\
	&= -\frac{1}{{N}}\int_{0}^{\infty}d\tau_1\int_{0}^{\infty}d\tau_2 [R^{(1)}(\tau_1,\tau_2)+R^{(1)}(\tau_1,\tau_2)^{*} - R^{(2)}(\tau_1,\tau_2) - R^{(2)}(\tau_1,\tau_2)^{*}]e^{i\omega_1 \tau_1}e^{i\omega_2 \tau_2}\\
\end{split}
\end{equation}

Performing the Fourier transformations and calling $E_0$ the ground state energy, we find, for the  $R_1 = R^{(1)} +R^{(1)*}$ contributions,

\begin{equation}
\begin{split}
\label{eq:R1}
\chi^{abc}_{R_1} (\omega_1,\omega_2) &= -\frac{1}{{N}}\sum_{PQ} \sum_{klm}  \bra{0} \sigma_k^a \ket{Q} \bra{Q}\sigma_l^b\ket{P}\bra{P} \sigma_m^{c} \ket{0} \frac{i}{\omega_1-(E_P-E_0)+i\Gamma} \frac{i}{\omega_2-(E_Q-E_0)+i\Gamma}\\
    &~~ -\frac{1}{{N}}\sum_{PQ} \sum_{klm}  (\bra{0} \sigma_k^a \ket{Q} \bra{Q}\sigma_l^b\ket{P}\bra{P} \sigma_m^{c} \ket{0})^{*} \frac{i}{\omega_1+(E_P-E_0)+i\Gamma} \frac{i}{\omega_2+(E_Q-E_0)+i\Gamma}\\
    & = -\frac{1}{{N}}\sum_{PQ} \sum_{klm}  \bra{0} \sigma_k^a \ket{Q} \bra{Q}\sigma_l^b\ket{P}\bra{P} \sigma_m^{c} \ket{0} g(\omega_1-(E_P-E_0)) g(\omega_2-(E_Q-E_0))\\
    &~~ -\frac{1}{{N}}\sum_{PQ} \sum_{klm}  (\bra{0} \sigma_k^a \ket{Q} \bra{Q}\sigma_l^b\ket{P}\bra{P} \sigma_m^{c} \ket{0})^{*}  g(\omega_1 + (E_P-E_0)) g(\omega_2 + (E_Q-E_0)),\\
\end{split}
\end{equation}
Proceeding along the same lines for $R_{2}$,
\begin{equation}
\begin{split}
\label{eq:R2}
\chi^{abc}_{R_2}(\omega_1,\omega_2) &= +\frac{1}{{N}}\sum_{PQ} \sum_{klm}  \bra{0} \sigma_l^b \ket{Q} \bra{Q} \sigma_k^a \ket{P}\bra{P} \sigma_m^{c} \ket{0} \frac{i}{\omega_1-(E_P-E_0)+i\Gamma} \frac{i}{\omega_2-(E_P-E_Q)+i\Gamma} \\
&~~+\frac{1}{{N}}\sum_{PQ} \sum_{klm}  (\bra{0} \sigma_l^b \ket{Q} \bra{Q} \sigma_k^a \ket{P}\bra{P} \sigma_m^{c} \ket{0})^{*} \frac{i}{\omega_1+(E_P-E_0)+i\Gamma} \frac{i}{\omega_2+(E_P-E_Q)+i\Gamma}\\
&= +\frac{1}{{N}}\sum_{PQ} \sum_{klm}  \bra{0} \sigma_l^b \ket{Q} \bra{Q} \sigma_k^a \ket{P}\bra{P} \sigma_m^{c} \ket{0} g(\omega_1-(E_P-E_0))g(\omega_2-(E_P-E_Q)) \\
&~~+\frac{1}{{N}}\sum_{PQ} \sum_{klm}  (\bra{0} \sigma_l^b \ket{Q} \bra{Q} \sigma_k^a \ket{P}\bra{P} \sigma_m^{c} \ket{0})^{*} g(\omega_1+(E_P-E_0))g(\omega_2+(E_P-E_Q)).\\
\end{split}
\end{equation}

Putting them together,
\begin{equation}
\chi^{abc}(\omega_1,\omega_2) = \chi^{abc}_{R_1} (\omega_1,\omega_2) +\chi^{abc}_{R_2}(\omega_1,\omega_2).
\end{equation}

This is the expression analyzed in the main text.

\begin{figure}
    \centering
    \includegraphics[width=0.95\linewidth]{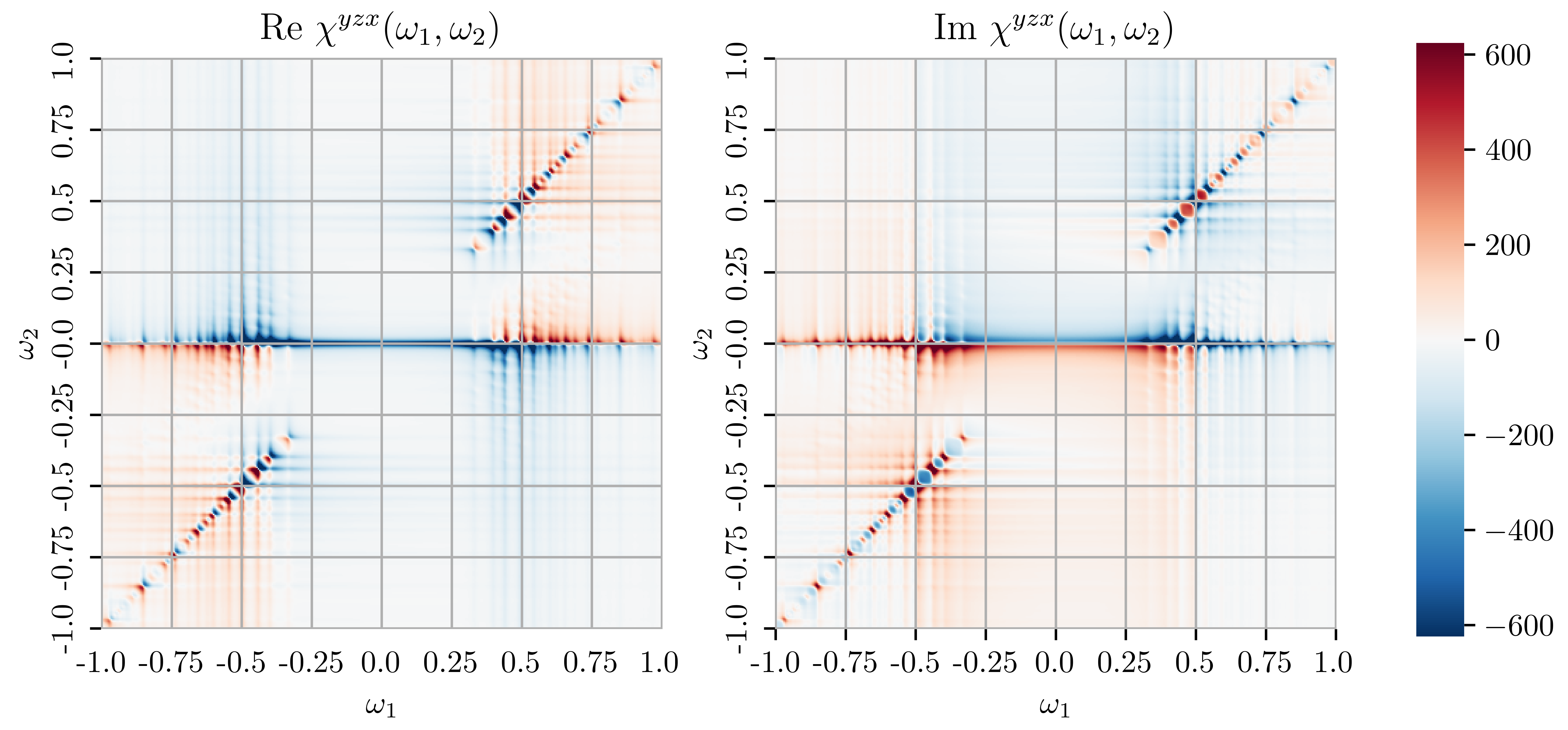}
    \caption{Total contribution of $R_1$ and $R_2$. $L=100, J_i = 1.0$. }
    \label{fig:total}
\end{figure}

\begin{figure}
    \centering
    \includegraphics[width=0.95\linewidth]{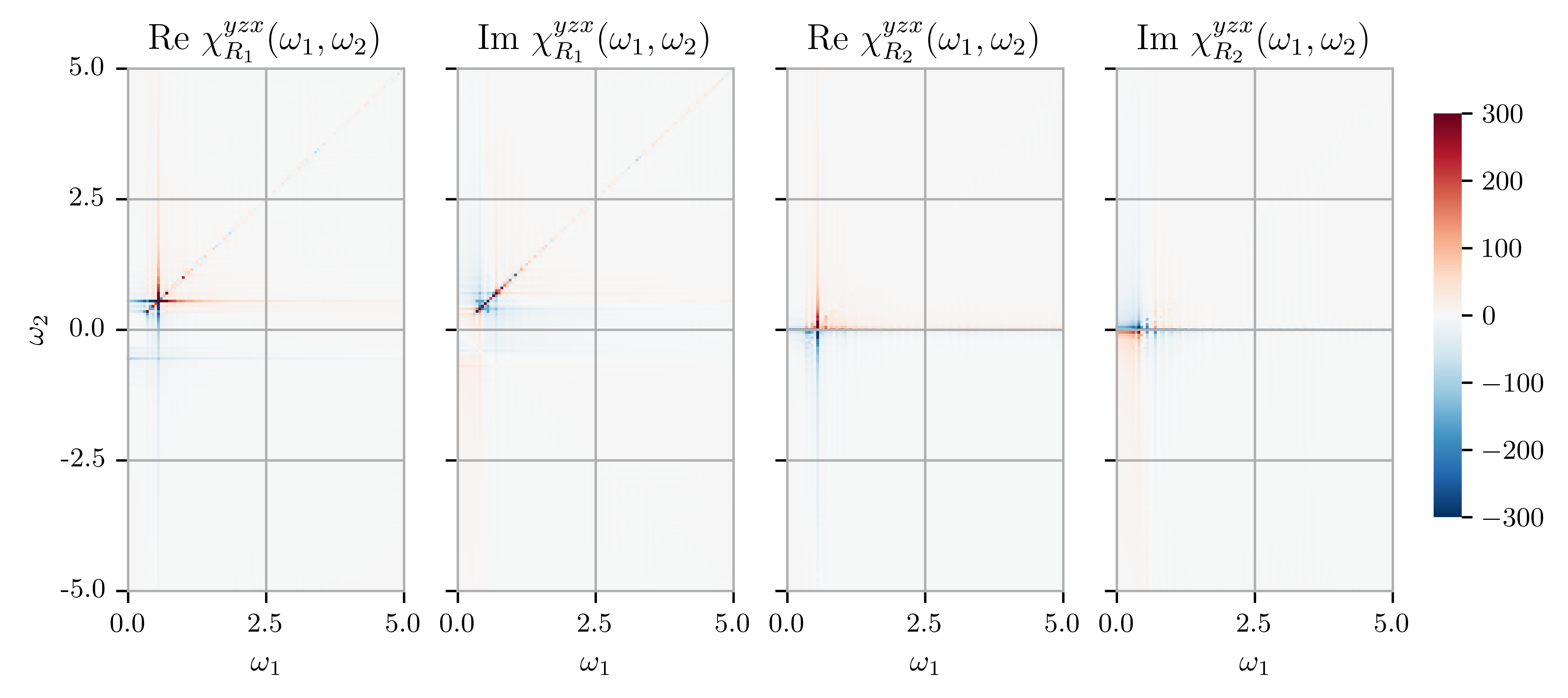}
    \caption{Real and imaginary part of $\chi_{R_1}^{yzx}$ and $\chi_{R_1}^{yzx}$ with large frequency range. $L=80, J_i = 1.0$. We see the signals mostly appear in the low-frequency region.}
    \label{fig:R1R2_large_freq}
\end{figure}

\section{Details of the solution of the Hamiltonian and the matrix elements of the second-order correlation functions \label{sec:mat_elem}}
In this Section, we give further details on how to solve the Hamiltonian and how to calculate the matrix elements entering the response functions. 

\subsection{Diagonalization of the Hamiltonian and gauge structure}

Following Kitaev's solution~\cite{kitaevAnyonsExactlySolved2006}, we introduce the complex bond fermions
\begin{equation}
\chi_{\langle j k\rangle}^{a}=\frac{1}{2}\left(b_{j}^{a}+i b_{k}^{a}\right),\quad (\chi_{\langle j k\rangle}^{a})^{\dagger}=\frac{1}{2}\left(b_{j}^{a}-i b_{k}^{a}\right).
\end{equation}
For simplicity we denote $\chi_{\langle j k\rangle}^{a}$ as $\chi_{jk}$ from now on. Since $u_{jk} = 2\chi^{\dag}_{jk}\chi_{jk} - 1$, the $\chi$ fermion describes the occupation of a bond $\langle j,k\rangle$ along the direction $a$. A bond $\langle j,k\rangle$ is occupied by $\chi$ fermion if $u_{jk}=+1$ or empty if $u_{jk}=-1$. 
Together with the fermionic mapping, we express the spin operators as 

\begin{equation}
\begin{split}
    \sigma_{i}^{a}&=i \left(\chi_{i j}+\chi_{i j}^{\dagger}\right)c_{Ai}\\
    \sigma_{j}^{a}&=\left(\chi_{i j}-\chi_ {i j}^{\dagger}\right) c_{Bj}
\end{split}
\end{equation}
The action of a spin operator can be viewed as applying a matter Majorana fermion and flipping the value of the bond $u_{jk}$ variable. The latter action corresponds to introducing two fluxes in plaquettes adjacent to the $a$-type bond.

The problem becomes to write the eigenstates of the matter fermions moving in a background of fluxes characterized by the set $\{u\}$. The ground state lies in the flux-free sector for large enough systems with spatial translational invariance~\cite{zschockePhysicalStatesFinitesize2015}. The conventional choice for the ground state gauge is $u=+1$. Obviously, all gauge configurations leading to the same flux sector will be equivalent and gives the same fermion energies.

The singular-value-decomposition (SVD) of the matter Hamiltonian leads to a natural definition of complex matter fermions~\cite{zschockePhysicalStatesFinitesize2015}, 
\begin{equation}
\label{svd_0}
\begin{aligned}
H_{\{u\}} &=\frac{i}{2}\left(\begin{array}{cc}
\mathbf{c}_{A}^{T} & \mathbf{c}_{B}^{T}
\end{array}\right)\left(\begin{array}{cc}
0 & M \\
-M^{T} & 0
\end{array}\right)\left(\begin{array}{c}
\mathbf{c}_{A} \\
\mathbf{c}_{B}
\end{array}\right)=\frac{i}{2}\left(\begin{array}{cc}
\mathbf{c}_{A}^{T} & \mathbf{c}_{B}^{T}
\end{array}\right)\left(\begin{array}{cc}
0 & U S V^{T} \\
-V S U^{T} & 0
\end{array}\right)\left(\begin{array}{c}
\mathbf{c}_{A} \\
\mathbf{c}_{B}
\end{array}\right) \\
&=\frac{i}{2}\left(\begin{array}{cc}
\mathbf{c}_{A}^{T} & \mathbf{c}_{B}^{T}
\end{array}\right)\left(\begin{array}{cc}
U & 0 \\
0 & V
\end{array}\right)\left(\begin{array}{cc}
0 & S \\
-S & 0
\end{array}\right)\left(\begin{array}{ll}
U & 0 \\
0 & V
\end{array}\right)^{T}\left(\begin{array}{c}
\mathbf{c}_{A} \\
\mathbf{c}_{B}
\end{array}\right) \\
&=\frac{i}{2}\left(\begin{array}{ll}
\left(\mathbf{e}^{\prime}\right)^{T} & \left(\mathbf{e}^{\prime \prime}\right)^{T}
\end{array}\right)\left(\begin{array}{cc}
0 & S \\
-S & 0
\end{array}\right)\left(\begin{array}{l}
\mathbf{e}^{\prime} \\
\mathbf{e}^{\prime \prime}
\end{array}\right)= i \sum_{m=0}^{N -1} \varepsilon_{m} e_{m}^{\prime} e_{m}^{\prime \prime}=\sum_{m=0}^{N-1} 2\varepsilon_{m}\left(a_{m}^{\dagger} a_{m}-\frac{1}{2}\right),
\end{aligned}
\end{equation}
where $e_{m}$ are Majorana modes and $a_m$ are matter fermion excitations with $a_m = \frac{1}{2}(e_{m}^{\prime} + i e_{m}^{\prime \prime})$. The vector $\mathbf{c}_{A(B)}$ is of length $N$ ($N\equiv L^2$ is the number of unit cells). We call the ground state complex matter excitation $a_m^{\dag}$, related to the matter Majoranas $c$ by 

\begin{equation}
\begin{split}
\label{c_to_a}
c_{Ai} &= \sum_{m}(U_0)_{im} (a_m^{\dag} + a_m),\\
c_{Bj} &= \sum_{m}(iV_0)_{jm} (a_m^{\dag} - a_m).
\end{split}
\end{equation}

 As a consequence of enlarging the Hilbert space, not all possible occupations of $a$ are physically acceptable~\cite{kitaevAnyonsExactlySolved2006}. 
 As mentioned in the main text, the Majorana operators act on the extended 4-dimensional Fock space $\tilde{\mathcal{M}}$, whereas the physical Hilbert space $\mathcal{M}$ of a spin is a subspace of $\tilde{\mathcal{M}}$ defined by~\cite{kitaevAnyonsExactlySolved2006, zschockePhysicalStatesFinitesize2015, knolleDynamicsTwoDimensionalQuantum2014}
\begin{equation}
|\xi\rangle \in \mathcal{M} \iff  D_j |\xi\rangle=|\xi\rangle  \quad \forall j, \quad D_j=b_j^x b_j^y b_j^z c_j.
\end{equation}
This constraint also ensures the Majorana representation of the spins satisfies the $SU(2)$ algebra. A state $\ket{\Phi}$ is physical if $\mathcal{P}\ket{\Phi} = \ket{\Phi}$, where the projection operator $\ket{P}$ is~\cite{zschockePhysicalStatesFinitesize2015} 
\begin{equation}
\begin{split}
\mathcal{P} &= \prod_{i=1}^{2 N}\left(\frac{1+D_{i}}{2}\right)=\frac{1}{2^{2 N}} \sum_{\{j\}} \prod_{i \in\{j\}} D_{i}\\
	&= \left(\frac{1}{2^{2 N-1}} \sum_{\{j\}'} \prod_{i \in\{j\}'} D_{i}\right) \cdot\left(\frac{1+\prod_{i=1}^{2 N} D_{i}}{2}\right)\\
	&= \mathcal{S} \cdot \mathcal{P}_{0}
\end{split}
\end{equation}
where $\{j\}$ runs over all possible subsets of site index set $\lambda$, while $\{j\}'$ is restricted to half of it (meaning $\{j\}'$ will not be $\{i\}$ and the complementary set $\lambda - \{i\}$ at the same time; these $2^{2N-1}$ terms give all the inequivalent transformations).
Here $\mathcal{S}$ symmetrically sums over physically equivalent eigenstates and $\mathcal{P}_0$ projects out the unphysical states.
$D_i$, the gauge transformation operator acting on-site $i$, can be rewritten in terms of complec fermions as
\begin{equation}
\begin{aligned}
&D_{iA} = \left[\chi_{ij}^{x}+\left(\chi_{ij}^{x}\right)^{\dagger}\right]\left[\chi_{ij}^{y} + \left(\chi_{ij}^{y}\right)^{\dagger}\right]\left[\chi_{ij}^{z} + \left(\chi_{ij}^{z}\right)^{\dagger}\right] c_i, \\
&D_{jB} = i\left[\chi_{ij}^{x}- \left(\chi_{ij}^{x}\right)^{\dagger}\right]\left[\chi_{ij}^{y} - \left(\chi_{ij}^{y}\right)^{\dagger}\right]\left[\chi_{ij}^{z} - \left(\chi_{ij}^{z}\right)^{\dagger}\right] c_j,
\end{aligned}
\end{equation}

\subsection{The matrix elements}

As discussed in the main text, purely from the flux constraints, the only possible non-vanishing polarization combinations are $a, b, c = x, y, z$ and their permutations. Below we show an example where $a=y, b=z, c=x$. The local structure of the flux operations also simplifies the summation over sites $\sum_{k,l,m}$, as demonstrated below. 

As an example, we show how to simplify the first line in $\chi^{yzx}_{R_1}$, Eq.~\eqref{eq:R1} and compute its matrix elements. The other matrix elements in Eq.~\eqref{eq:R1} and Eq.~\eqref{eq:R2} can be computed similarly. The first line of $\chi^{yzx}_{R_1}$ is called $\chi^{yzx}_{R_1,1} (\omega_1,\omega_2)$ and given by
\begin{equation}
    \chi^{yzx}_{R_1,1} (\omega_1,\omega_2) = -\frac{1}{{N}}\sum_{PQ} \sum_{klm}  \bra{0} \sigma_k^y \ket{Q} \bra{Q}\sigma_l^z\ket{P}\bra{P} \sigma_m^{x} \ket{0} g(\omega_1-(E_P-E_0)) g(\omega_2-(E_Q-E_0)).
\end{equation}

The strategy is to fix the site $m$ and look at all possible neighboring sites contributing to the sum.  We label the unit cell associated with site $m$ as $\nu$ and call the two sublattices $A$ and $B$. To understand which are the unit cells neighboring $\nu$ that contribute to the sum, the easiest is to draw the lattice and its connection, as shown in Fig.~\ref{fig:unit_cell}. The expression becomes 
\begin{equation}
\label{eq:R11}
\begin{split}
\chi^{yzx}_{R_1,1} (\omega_1,\omega_2) &= -\frac{1}{{N}}\sum_{PQ} \sum_{\nu}  [\bra{0} \sigma_{\nu A}^y + \sigma_{\nu-a_2 B}^y \ket{Q} \bra{Q}\sigma_{\nu A}^z+\sigma_{\nu B}^z\ket{P}\bra{P} \sigma_{\nu A}^{x} +\sigma_{\nu-a_1 B}^{x} \ket{0}\\
	&~~~~~~~~~~~~~~~~~+\bra{0} \sigma_{\nu B}^y + \sigma_{\nu+a_2 A}^y \ket{Q} \bra{Q}\sigma_{\nu A}^z+\sigma_{\nu B}^z\ket{P}\bra{P} \sigma_{\nu B}^{x} +\sigma_{\nu+a_1 A}^{x} \ket{0}]\\
	&~~~~~~~~~~~~~~~~~~~~~~~~~~~~~~~~~~~~~~~~~~~~~~~~~~~~~~~~\times g(\omega_1-(E_P-E_0)) g(\omega_2-(E_Q-E_0)),
\end{split}
\end{equation}
where $\nu \pm a_i$ labels the neighboring unit cells in $\pm\bm{a_i}$ directions with $\bm{a_i}$ the basis vectors for honeycomb lattice. This important simplification reduced the sum over three indices to a sum over a single index and is a clear consequence of the fluxes getting created locally. By translation symmetry, the last sum can also be reduced to the structure shown in Fig.~\ref{fig:unit_cell}, which means that all we have to calculate involves a fixed value of $\nu$ and multiplying by the number of unit cells. In fact, we can compute the lower four sites (Benz star) and multiply the result by two, given the symmetry around the $z$ vertical bond. The factor of $N$ unit cells is canceled by the $1/N$ factor in the definition of the response function.
\begin{figure}[H]
\centering
\includegraphics[width=0.25\textwidth]{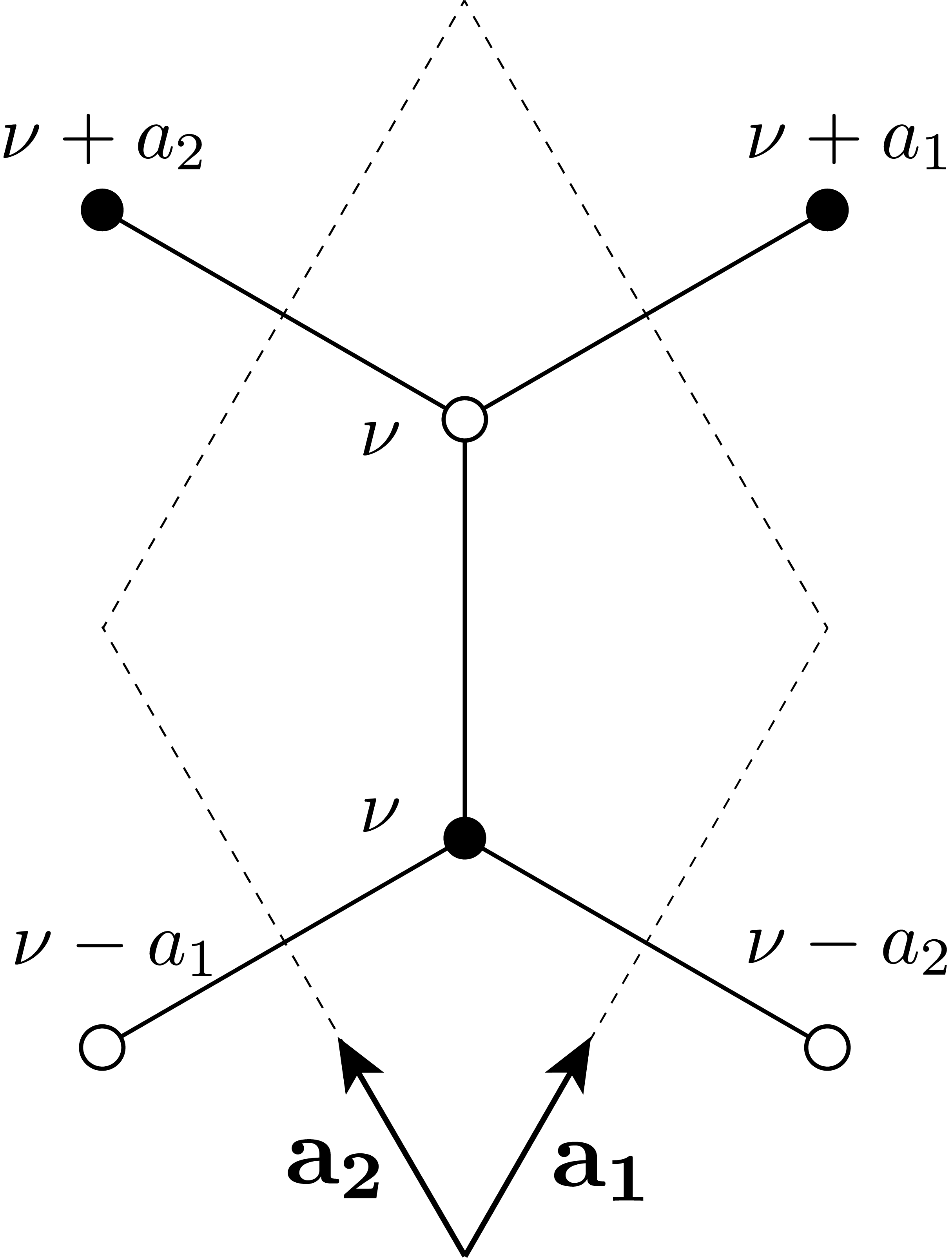}
\caption{Summation over unit cell index $\nu$ involved in Eq.~\eqref{eq:R11}. $\bm{a_1}$ and $\bm{a_2}$ are base vectors. Solid (hollow) dots represent sites belonging to $A(B)$ sublattice. The sites $(\nu, A)$ and $(\nu, B)$ are connected by a $z$ bond.}
\label{fig:unit_cell}
\end{figure}

We consider the intermediate states $\ket{P}$ and $\ket{Q}$ to have two fluxes, as the ground state is flux-free and each spin operator adds two fluxes to the system. As for the matter sector, we consider up to one matter fermion in these intermediate states. It has been argued that considering one particle in the matter sector provides a good approximation to capture the main physics~\cite{choiTheoryTwoDimensionalNonlinear2020}.

The complex matter excitation in the 2-flux sector $b_{\lambda}^{\dag} (\equiv a^{\mathbf{u}_\text{2flux}})$ is related to the complex matter excitation $a_m^{\dag}(\equiv a^{\mathbf{u}_\text{flux-free}})$ in the flux-free sector~\cite{blaizotQuantumTheoryFinite1985} by 
\begin{equation}
\begin{split}
b_{\lambda} &= \sum_{m}(X_0^2)^*_{\lambda m} a_m + (Y_0^2)^*_{\lambda m} a_m^{\dag},\\
b_{\lambda}^{\dag} &= \sum_{m}(X_0^2)_{\lambda m} a_m^{\dag} + (Y_0^2)_{\lambda m} a_m ,
\end{split}
\end{equation}
where 
\begin{equation}
\begin{split}
X_0^{2*} &= \frac{1}{2}(U_2^{\dag}U_0 + V_2^{\dag}V_0),\\
Y_0^{2*} &= \frac{1}{2}(U_2^{\dag}U_0 - V_2^{\dag}V_0),
\end{split}
\end{equation}
with $U_{0(2)},V_{0(2)}$ the orthogonal matrices given by the SVD transformation, Eq.~\eqref{svd_0}.

We also relate the matter vacuum of the 2-flux state $\ket{M_0^{2}}$ to that of ground state $\ket{M_0}$~\cite{knolleDynamicsQuantumSpin2016}, 
\begin{equation}
\ket{M_0^2} = |\det X_0^2|^{\frac{1}{2}} e^{-\frac{1}{2}\sum a_m^{\dag}F_{mn}a_n^{\dag}}\ket{M_0},
\end{equation}
with
\begin{equation}
\begin{split}
F &= (X_0^{2*})^{-1}Y_0^{2*}, \quad (F^{T}=-F).
\end{split}
\end{equation}
 To compute the matrix elements, we use the following relations derived using Wick's theorem, 
\begin{equation}
\label{elements1}
\begin{split}
\bra{M_0} c_{Ai} b^{\dag}_{\lambda} \ket{M_0^2} &= |\det X_0^2|^{\frac{1}{2}} [U_0(X_0^{2})^{-1}]_{i \lambda}\\
\bra{M_0^2} b_{\lambda} c_{Bj}  \ket{M_0} &= |\det X_0^2|^{\frac{1}{2}} [iV_0(X_0^{2})^{-1}]_{j \lambda}\\
\bra{M_0}c_{Bj} b^{\dag}_{\lambda} \ket{M_0^2} &= |\det X_0^2|^{\frac{1}{2}} [-iV_0(X_0^{2})^{-1}]_{j \lambda}\\
\bra{M_0^2} b_{\lambda} c_{Ai}\ket{M_0} &= |\det X_0^2|^{\frac{1}{2}} [U_0(X_0^{2})^{-1}]_{i \lambda}.
\end{split}
\end{equation}
We devote particular attention to how to compute the middle matrix element $\bra{P}\mathcal{P} \sigma^z \mathcal{P}\ket{Q}$. The states $\ket{P}$ and $\ket{Q}$ have the same number of bond fermions and of matter excitations. The operator $\sigma_z$ in the middle changes the bond and matter fermion number by one and, therefore, without the projector operators, this matrix element would vanish. The projection operator plays, therefore, an important role in making this element finite. Computing the matrix element explicitly for unit cell $\nu = 0$, and sublattice site $A$, we find
\begin{equation}
\label{mid_element}
\begin{split}
\bra{Q}\mathcal{P} \sigma^z_{0A} \mathcal{P}\ket{P} &= \bra{Q}  [i(\chi_{0A0B}^{z} + \chi_{0A0B}^{z\dag})c_{0A}] \mathcal{P}\ket{P}\\
			&= \bra{M_0^{2'}} d_\mu \bra{F_{GS}} \chi_{0A,L(L-1)B}^{y\dag} [i(\chi_{0A0B}^{z} + \chi_{0A0B}^{z\dag})c_{0A}] (1+\sum_{j} D_{j}+\sum_{j<k} D_{j}D_{k}+\cdots ) \chi_{0A,LB}^{x}\ket{F_{GS}} b^{\dag}_{\lambda}\ket{M_0^2}\\
			&= \bra{M_0^{2'}} d_\mu \bra{F_{GS}} \chi_{0A,L(L-1)B}^{y\dag} [i(\chi_{0A0B}^{z} + \chi_{0A0B}^{z\dag})c_{0A}] D_{0A} \chi_{0A,LB}^{x}\ket{F_{GS}} b^{\dag}_{\lambda}\ket{M_0^2}\\
			&= i \bra{M_0^{2'}} d_\mu \bra{F_{GS}} \chi_{0A,L(L-1)B}^{y\dag} \chi_{0A0B}^{z\dag} c_{0A}   (\chi_{0A,LB}^{x\dag}\chi_{0A,L(L-1)B}^{y}\chi_{0A0B}^{z} c_{0A}) \chi_{0A,LB}^{x}\ket{F_{GS}} b^{\dag}_{\lambda}\ket{M_0^2}\\
			& = i\bra{M_0^{2'}} d_\mu b^{\dag}_{\lambda}\ket{M_0^2}\\
			& = i\sqrt{|\det X_P^Q|} [X_P^Q - Y_P^Q F_P^Q]_{\mu\lambda},
\end{split}
\end{equation}
where we named the matter excitations in $\ket{Q}$ and $\ket{P}$ $d$ and $b$, respectively, and the ground state gauge choice is $\ket{F_{GS}} = \chi_{1A,1B}^{z}\chi_{2A,0B}^{y}\chi_{3A,2B}^{x}\ket{F_0}$.
Similarly, we derive the matrix element for the spin operator located at the $B$ site of unit cell $\nu=0$,
\begin{equation}
\label{mid_element_B}
\begin{split}
\bra{Q}\mathcal{P} \sigma^z_{0B} \mathcal{P}\ket{P} &= \bra{Q}  [(\chi_{0A0B}^{z} - \chi_{0A0B}^{z\dag})c_{0B}] \mathcal{P}\ket{P}\\
			&= \bra{M_0^{2'}} d_\mu \bra{F_{GS}} \chi_{0A,L(L-1)B}^{y\dag} [(\chi_{0A0B}^{z} - \chi_{0A0B}^{z\dag})c_{0B}] (1+\sum_{j} D_{j}+\sum_{j<k} D_{j}D_{k}+\cdots ) \chi_{0A,LB}^{x}\ket{F_{GS}} b^{\dag}_{\lambda}\ket{M_0^2}\\
			&= \bra{M_0^{2'}} d_\mu \bra{F_{GS}} \chi_{0A,L(L-1)B}^{y\dag} [(\chi_{0A0B}^{z} - \chi_{0A0B}^{z\dag})c_{0B}] D_{0A} \chi_{0A,LB}^{x}\ket{F_{GS}} b^{\dag}_{\lambda}\ket{M_0^2}\\
			&= - \bra{M_0^{2'}} d_\mu \bra{F_{GS}} \chi_{0A,L(L-1)B}^{y\dag} \chi_{0A0B}^{z\dag} c_{0B}   (\chi_{0A,LB}^{x\dag}\chi_{0A,L(L-1)B}^{y}\chi_{0A0B}^{z} c_{0A}) \chi_{0A,LB}^{x}\ket{F_{GS}} b^{\dag}_{\lambda}\ket{M_0^2}\\
			& = \bra{M_0^{2'}} d_\mu c_{0A}c_{0B} b^{\dag}_{\lambda}\ket{M_0^2}\\
			& = i\sqrt{|\det X_P^Q|} [(X_P^Q)_{\mu \lambda} (U_P V^T_P)_{00} - (X_P U_P^T)_{\mu 0}(V_P)_{0\lambda} - (X_P V_P^T)_{\mu 0} (U_P)_{0\lambda}\\
			&~~~~~~~~~~~~~~~~ - (X_P U_P^T)_{\mu 0} (V_P F_P)_{0\lambda} + (X_P V_P^T)_{\mu 0} (U_P F_P)_{0\lambda}-(X_P)_{\mu\lambda} (U_P F_P V_P^T)_{00} \\
			&~~~~~~~~~~~~~~~~ - (Y_P F_P)_{\mu\lambda}(U_P V_P^T)_{00} + (Y_P F_P V_P^T)_{\mu 0}(U_P)_{0\lambda} + (Y_P F_P U_P^T)_{\mu 0} (V_P)_{0\lambda}\\
			&~~~~~~~~~~~~~~~~ + (Y_P F_P)_{\mu\lambda}(U_P F_P V_P^T)_{00} + (Y_P F_P^T V_P^T)_{\mu 0}(U_P F_P)_{0\lambda} + (Y_P F_P U_P^T)_{\mu 0}(V_P F_P)_{0\lambda}]
\end{split}
\end{equation}
\section{Contrasting the diagonal and off-diagonal peaks \label{sec:off_diag}}
In this Section, we show a way to understand the difference between the diagonal and off-diagonal peaks by looking at the ratio of the nonlinear response function that we calculated and the linear response ones.

For linear response, the linear spin-spin correlation function~\cite{zschockePhysicalStatesFinitesize2015} is often studied,
\begin{equation}
S_{i j}^{\alpha \beta}(t)=\bra{0}\hat{\sigma}_{i}^{\alpha}(t) \hat{\sigma}_{j}^{\beta}(0)\ket{0}
\end{equation}
For example the $xx$ components of optical ($q=0$) response in frequency space is
\begin{equation}
\begin{split}
S^{xx}(\omega) &= \frac{1}{N} \sum_{\langle i,j\rangle = x} S^{xx}_{AiAi}(\omega) + S^{xx}_{AiBj}(\omega) + S^{xx}_{BjAi}(\omega) + S^{xx}_{BjBj}(\omega)\\
&= 2\pi \sum_{\lambda} |\bra{P_\lambda} \sigma_{0A}^x +\sigma_{L-1B}^x \ket{0}|^2 \delta	[\omega-(E_\lambda - E_0)],			
\end{split}
\end{equation}
where $\ket{P_\lambda}$ is the eigenstate of the Hamiltonian with energy $E_\lambda$, and we have used the translational symmetry to simplify the equation.
The contribution from a certain (non-degenerate) level is 
\begin{equation}
S^{xx}_{\lambda}(\omega) = 2\pi  |\bra{P_\lambda} \sigma_{0A}^x +\sigma_{L-1B}^x \ket{0}|^2 \delta	[\omega-(E_\lambda - E_0)]
\end{equation}
We show the linear response computed with the whole set of eigenstates and with only a few localized high IPR states in Fig~\ref{fig:linear_res}. The result shows that in linear response the signal is mostly contributed by those localized states.
\begin{figure*}[t!]
  \includegraphics[width=0.8\linewidth]{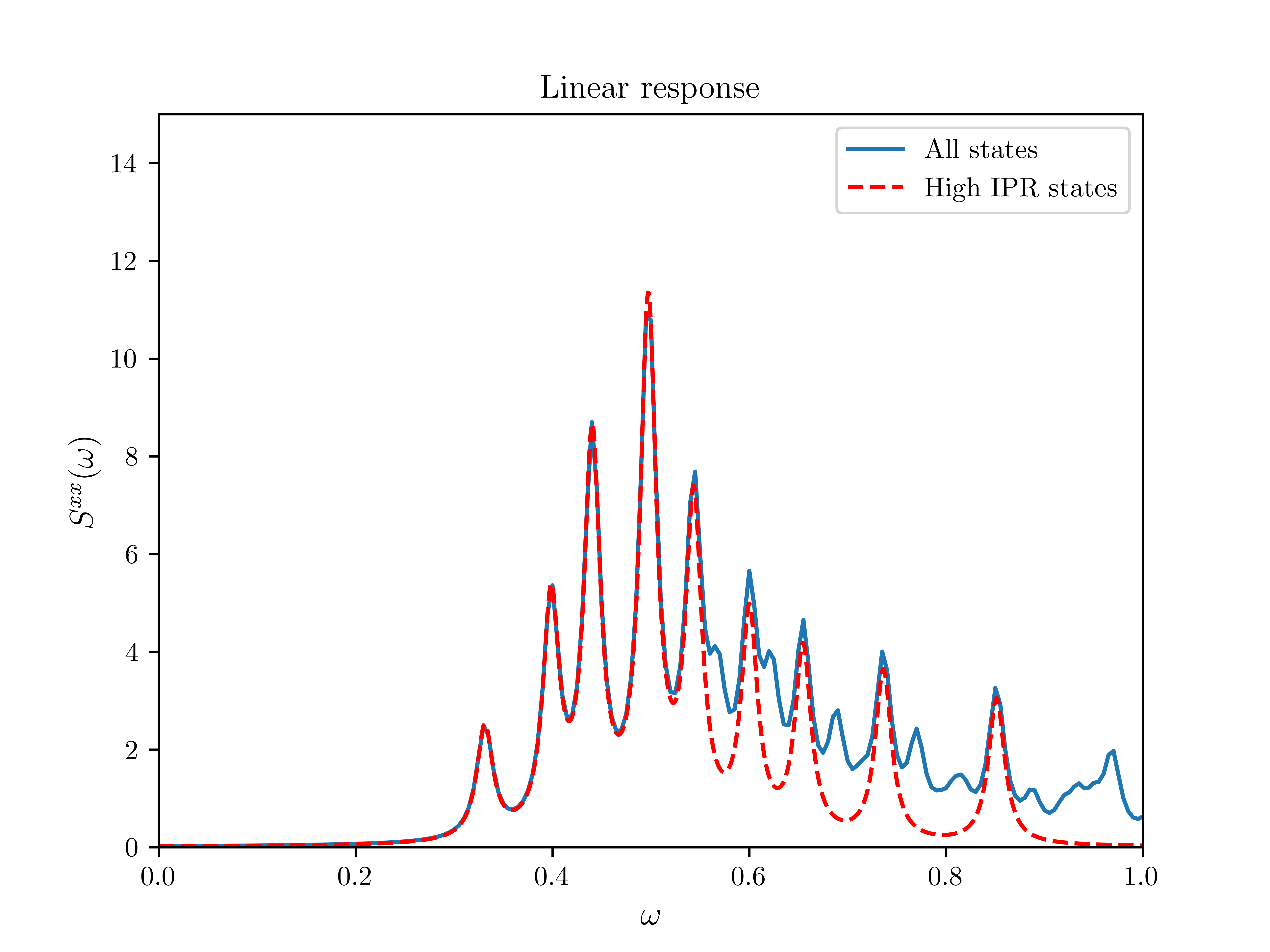}
  \caption{Linear response of Kitaev honeycomb model. $L=100$, $J_i=1$, $\Gamma=0.01$. The response is computed considering all the eigenstates (red line) and with only a few high IPR states.}
  \label{fig:linear_res}
\end{figure*}

As for the second order response we computed, we can extract the contribution of $\ket{P_\lambda}$, $\ket{Q_\mu}$ states from Eq.~\eqref{eq:R11}
\begin{equation}
\begin{split}
\chi^{y,z,x}_{R_1,1;\lambda\mu} (\omega_1,\omega_2) 
	&= - 2\bra{0} \sigma_{0 A}^y + \sigma_{L(L-1) B}^y \ket{Q_\mu} \bra{Q_\mu}\sigma_{0 A}^z+\sigma_{0 B}^z\ket{P_\lambda}\bra{P_\lambda} \sigma_{0 A}^{x} +\sigma_{L-1 B}^{x} \ket{0}\\
	&~~~~~~~~~~~~~~~~~~~~~~~~~~~~~~~~~~~~~~~~~~~~~~~~~~~~~~~~\times g(\omega_1-(E_{P_\lambda}-E_0)) g(\omega_2-(E_{Q_\mu}-E_0)),
\end{split}
\end{equation}
where we have used the $C_2$ symmetry to combine the matrix elements. For comparison, we take the ratio between the nonlinear and linear response contributions mentioned above. This ratio is proportional to the matrix element ratio,
\begin{equation}
\label{eqn:ratio}
\begin{split}
\left|\frac{\text{Im} \chi^{y,z,x}_{R_1,1;\lambda\mu}(\omega_1,\omega_2)}{S^{yy}_{\mu}(\omega_2)S^{xx}_{\lambda}(\omega_1)}\right| &=\frac{1}{2} \frac{|\bra{0} \sigma_{0 A}^y + \sigma_{L(L-1) B}^y \ket{Q_\mu} \bra{Q_\mu}\sigma_{0 A}^z+\sigma_{0 B}^z\ket{P_\lambda}\bra{P_\lambda} \sigma_{0 A}^{x} +\sigma_{L-1 B}^{x} \ket{0}|}{|\bra{Q_\lambda} \sigma_{0A}^y +\sigma_{L(L-1)B}^y \ket{0}|^2 |\bra{P_\lambda} \sigma_{0A}^x +\sigma_{L-1B}^x \ket{0}|^2}\\
&= \frac{1}{2} \left|\frac{\bra{Q_\mu}\sigma_{0 A}^z+\sigma_{0 B}^z\ket{P_\lambda}}{\bra{Q_\mu} \sigma_{0A}^y +\sigma_{L(L-1)B}^y \ket{0} \bra{P_\lambda} \sigma_{0A}^x +\sigma_{L-1B}^x \ket{0}}\right|
\end{split}
\end{equation}
We identify this quantity as a direct comparison between the nonlinear and linear response. The main factor here we see is the middle matrix element $\bra{Q_\mu}\sigma_{0 A}^z+\sigma_{0 B}^z\ket{P_\lambda}$, which is a new quantity that exists in the second order nonlinear response. This matrix element between $\ket{P}$ and $\ket{Q}$ is not present in linear or third-order responses. The previous matrix element $\bra{f}\sigma_l^{a}\ket{i}$ is between states sharing a different number of fluxes, while in our middle matrix element $\bra{P}\sigma_l^z\ket{Q}$, both $\ket{P}$ and $\ket{Q}$ have two fluxes. 
The $\lambda = \mu$ matrix element is, in fact, one order of magnitude larger than those with $\lambda \neq \mu$, which explains why the peaks along the diagonal are stronger than those that deviate from the diagonal in $R_1$ process.

\section{2D spectrum}
\begin{figure*}[t!]
  \includegraphics[width=\linewidth]{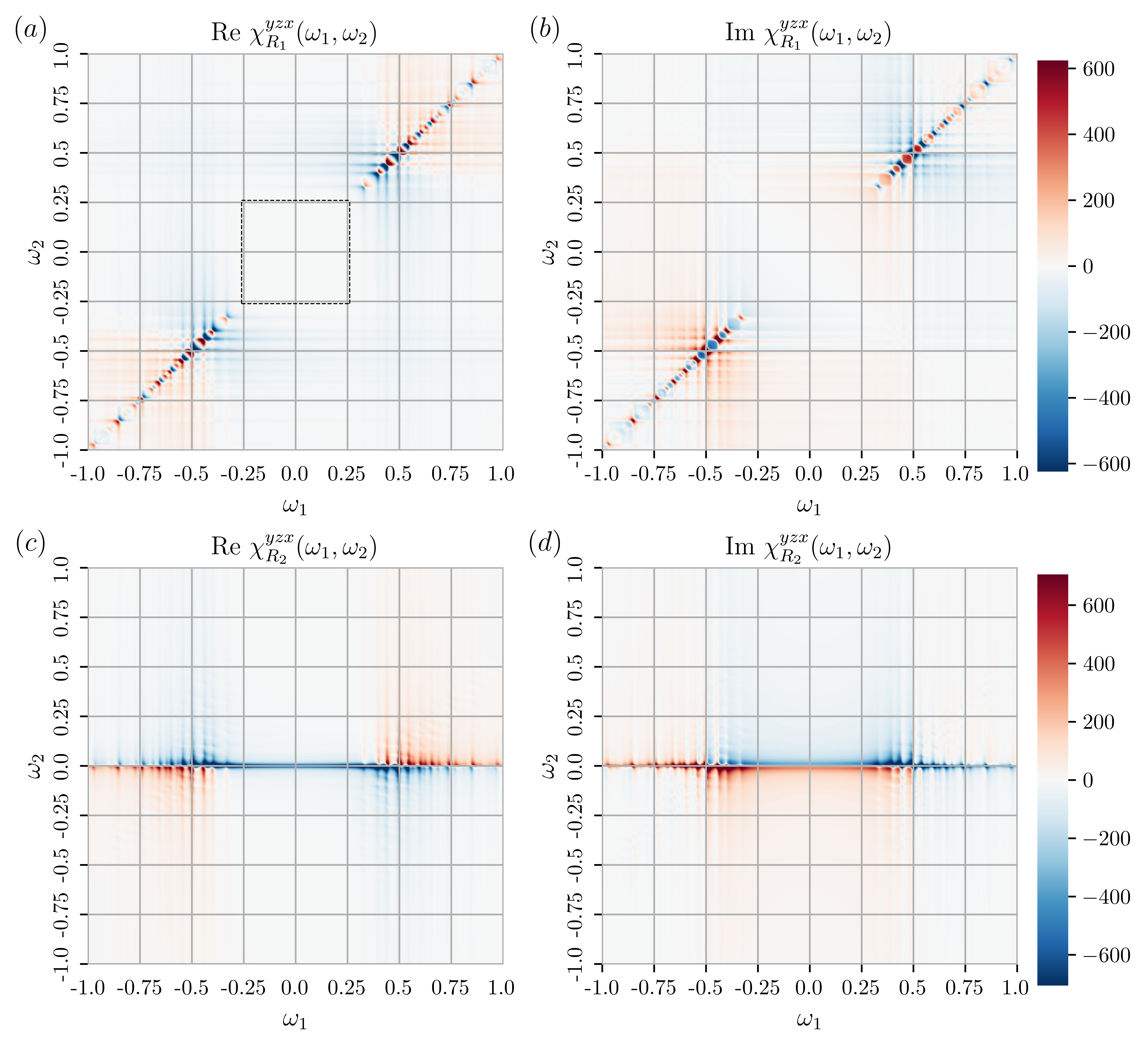}
  \caption{Contributions for the first nonlinear susceptibility $\chi^{yzx}$, $R_1$  ($a$ and $b$) and $R_{2}$ ($c$ and $d$), for the choice of parameters $L=100$, $J_i=1$ and $\Gamma=0.01$. ($a$ and $b$) Real and imaginary parts of $R_{1}(\omega_{1},\omega_{2})$. The black dashed box indicates the flux gap in the thermodynamic limit. The proximity to the diagonal is determined by how anisotropic the coupling constants are. The strong peak along the diagonal comes from the overlap of states with two neighboring fluxes and trapped Majorana matter fermions. ($c$ and $d$) Real and imaginary part of $R^{(2)}(\omega_{1},\omega_{2}))$. In this case, $\omega_{2}$ measures the energy difference between fluxes neighboring $x$ and $y$ bonds. In the isotropic case that we are considering, the peaks are along the $\omega_{2}=0$ line. The position of the peak at $\omega_{1}$ is the same as for $R_{1}$.}
  \label{fig:2dcs}
\end{figure*}
\section{Away from the isotropic point \label{sec:aniso}}

This Section considers the cases in which $J_{x}\ne J_{y}$. Below Fig.~\ref{fig:R1_aniso} shows the 2D spectrum of $R^{(1)}$ process, with ansotropy. Given $J_y>J_x$, and that in $R_1$ process $E_Q-E_0$ is probed by $\omega_2$, so the peaks now appear above $\omega_2=\omega_1$ diagonal.  

\begin{figure}[!tb]
    \centering
    \includegraphics{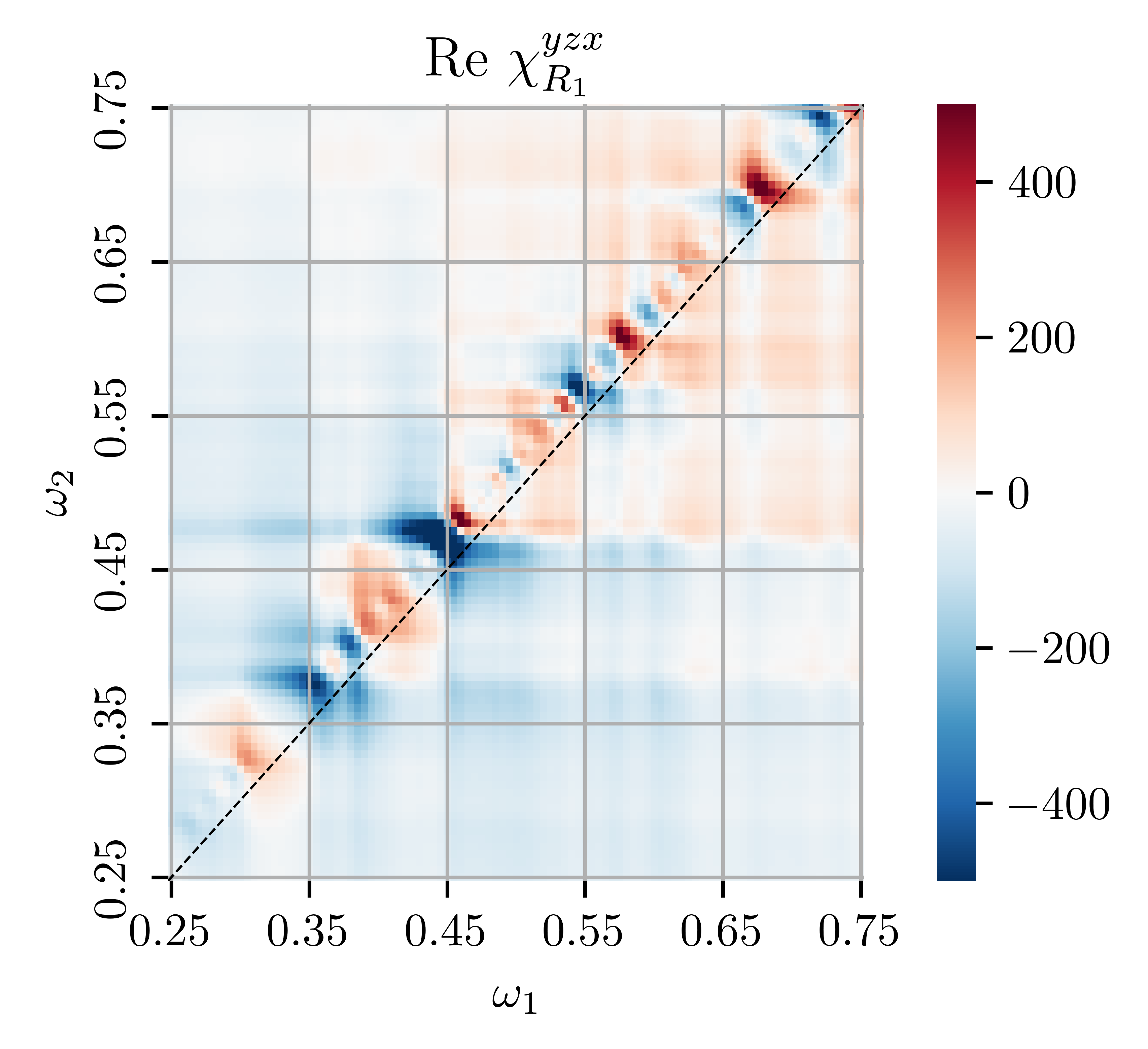}
    \caption{Anisotropic zoomed-in plot for $0.25<\omega_{1,2}<0.75$; $L=100, J_x=0.86, J_y=0.95, J_z=1.19$. The black dashed line indicates the diagonal $\omega_1=\omega_2$. }
    \label{fig:R1_aniso}
\end{figure}

\begin{figure}[!tb]
    \centering
    \includegraphics[width=\linewidth]{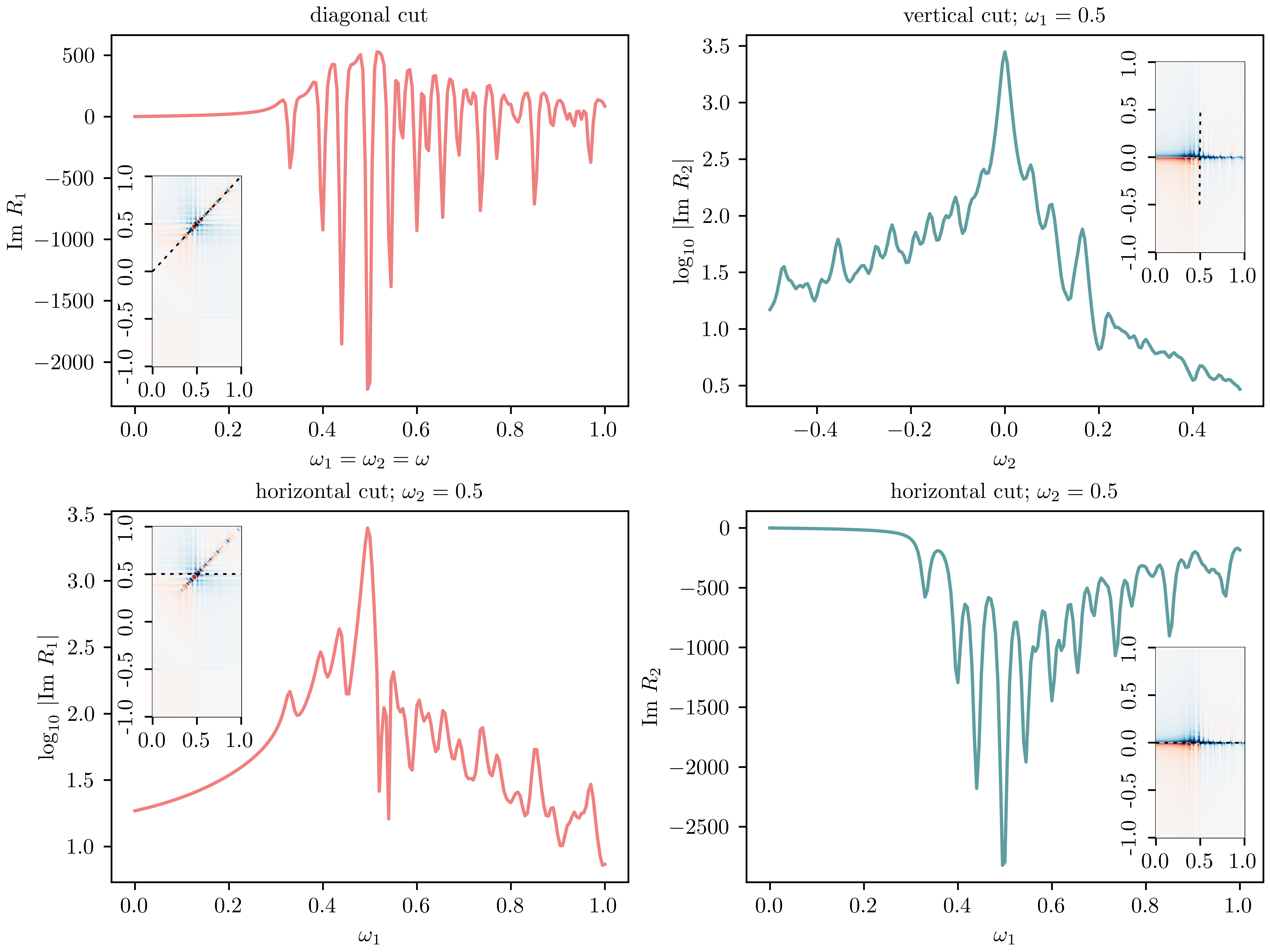}
    \caption{Cut plots for the 2D spectra.}
    \label{fig:four_cuts}
\end{figure}

\section{The inverse participation ratio of the matter states \label{sec:ipr}}
Here, we give further details on differentiating the localized and extended matter states according to their IPR. From Eq.~\eqref{c_to_a}, we express zero flux complex matter excitation in terms of real Majorana matter fermions as $a_m^{\dag} = \frac{1}{2} \sum (U^{T}_{m i} c_{A i} - i V^{T}_{m j} c_{B j})$. Thus for 0-flux matter state wave-function $\psi^{(0)}_{m} (\textbf{r}) = \bra{\textbf{r}}a^{\dag}_m\ket{0}$, we associate $(U_0^{T})_{m\nu}$ (or $(V_0^{T})_{m\nu}$) as the amplitude of applying Majorana $c$ fermion on A (or B) site in unit cell $\nu$. Generally, for the $l$-flux sector, we define the inverse participation ratio (IPR) of a real space wavefunction the according to $(U_l^{T})$ and $(V_l^{T})$ as 
\begin{equation}
    \text{IPR}(\psi^{(l)}_m) = \frac{\sum_{k}(U_l^{T})_{m,k}^4+(V_l^{T})_{m,k}^4}{\sum_{k}(U_l^{T})_{m,k}^2+(V_l^{T})_{m,k}^2},
\end{equation}
where $l$ labels what type of flux we have for state $\psi^{(l)}_m$. For example, for $R^{(1)}$ process the $\ket{P}$ states have $(l=2,x)$ while $\ket{Q}$ states have $(l=2,y)$.
\end{widetext}


\begin{thebibliography}{50}%
\makeatletter
\providecommand \@ifxundefined [1]{%
 \@ifx{#1\undefined}
}%
\providecommand \@ifnum [1]{%
 \ifnum #1\expandafter \@firstoftwo
 \else \expandafter \@secondoftwo
 \fi
}%
\providecommand \@ifx [1]{%
 \ifx #1\expandafter \@firstoftwo
 \else \expandafter \@secondoftwo
 \fi
}%
\providecommand \natexlab [1]{#1}%
\providecommand \enquote  [1]{``#1''}%
\providecommand \bibnamefont  [1]{#1}%
\providecommand \bibfnamefont [1]{#1}%
\providecommand \citenamefont [1]{#1}%
\providecommand \href@noop [0]{\@secondoftwo}%
\providecommand \href [0]{\begingroup \@sanitize@url \@href}%
\providecommand \@href[1]{\@@startlink{#1}\@@href}%
\providecommand \@@href[1]{\endgroup#1\@@endlink}%
\providecommand \@sanitize@url [0]{\catcode `\\12\catcode `\$12\catcode
  `\&12\catcode `\#12\catcode `\^12\catcode `\_12\catcode `\%12\relax}%
\providecommand \@@startlink[1]{}%
\providecommand \@@endlink[0]{}%
\providecommand \url  [0]{\begingroup\@sanitize@url \@url }%
\providecommand \@url [1]{\endgroup\@href {#1}{\urlprefix }}%
\providecommand \urlprefix  [0]{URL }%
\providecommand \Eprint [0]{\href }%
\providecommand \doibase [0]{https://doi.org/}%
\providecommand \selectlanguage [0]{\@gobble}%
\providecommand \bibinfo  [0]{\@secondoftwo}%
\providecommand \bibfield  [0]{\@secondoftwo}%
\providecommand \translation [1]{[#1]}%
\providecommand \BibitemOpen [0]{}%
\providecommand \bibitemStop [0]{}%
\providecommand \bibitemNoStop [0]{.\EOS\space}%
\providecommand \EOS [0]{\spacefactor3000\relax}%
\providecommand \BibitemShut  [1]{\csname bibitem#1\endcsname}%
\let\auto@bib@innerbib\@empty
\bibitem [{\citenamefont {Dressel}\ and\ \citenamefont
  {Gr{\"u}ner}(2002)}]{dresselElectrodynamicsSolidsOptical2002}%
  \BibitemOpen
  \bibfield  {author} {\bibinfo {author} {\bibfnamefont {M.}~\bibnamefont
  {Dressel}}\ and\ \bibinfo {author} {\bibfnamefont {G.}~\bibnamefont
  {Gr{\"u}ner}},\ }\href {https://doi.org/10.1017/CBO9780511606168} {\emph
  {\bibinfo {title} {Electrodynamics of {{Solids}}: {{Optical Properties}} of
  {{Electrons}} in {{Matter}}}}},\ \bibinfo {edition} {1st}\ ed.\ (\bibinfo
  {publisher} {{Cambridge University Press}},\ \bibinfo {year}
  {2002})\BibitemShut {NoStop}%
\bibitem [{\citenamefont {Basov}\ \emph {et~al.}(2011)\citenamefont {Basov},
  \citenamefont {Averitt}, \citenamefont {{van der Marel}}, \citenamefont
  {Dressel},\ and\ \citenamefont
  {Haule}}]{basovElectrodynamicsCorrelatedElectron2011}%
  \BibitemOpen
  \bibfield  {author} {\bibinfo {author} {\bibfnamefont {D.~N.}\ \bibnamefont
  {Basov}}, \bibinfo {author} {\bibfnamefont {R.~D.}\ \bibnamefont {Averitt}},
  \bibinfo {author} {\bibfnamefont {D.}~\bibnamefont {{van der Marel}}},
  \bibinfo {author} {\bibfnamefont {M.}~\bibnamefont {Dressel}},\ and\ \bibinfo
  {author} {\bibfnamefont {K.}~\bibnamefont {Haule}},\ }\href
  {https://doi.org/10.1103/RevModPhys.83.471} {\bibfield  {journal} {\bibinfo
  {journal} {Rev. Mod. Phys.}\ }\textbf {\bibinfo {volume} {83}},\ \bibinfo
  {pages} {471} (\bibinfo {year} {2011})}\BibitemShut {NoStop}%
\bibitem [{\citenamefont {Devereaux}\ and\ \citenamefont
  {Hackl}(2007)}]{devereauxInelasticLightScattering2007}%
  \BibitemOpen
  \bibfield  {author} {\bibinfo {author} {\bibfnamefont {T.~P.}\ \bibnamefont
  {Devereaux}}\ and\ \bibinfo {author} {\bibfnamefont {R.}~\bibnamefont
  {Hackl}},\ }\href {https://doi.org/10.1103/RevModPhys.79.175} {\bibfield
  {journal} {\bibinfo  {journal} {Rev. Mod. Phys.}\ }\textbf {\bibinfo {volume}
  {79}},\ \bibinfo {pages} {175} (\bibinfo {year} {2007})}\BibitemShut
  {NoStop}%
\bibitem [{\citenamefont {Sobota}\ \emph {et~al.}(2021)\citenamefont {Sobota},
  \citenamefont {He},\ and\ \citenamefont
  {Shen}}]{sobotaAngleresolvedPhotoemissionStudies2021}%
  \BibitemOpen
  \bibfield  {author} {\bibinfo {author} {\bibfnamefont {J.~A.}\ \bibnamefont
  {Sobota}}, \bibinfo {author} {\bibfnamefont {Y.}~\bibnamefont {He}},\ and\
  \bibinfo {author} {\bibfnamefont {Z.-X.}\ \bibnamefont {Shen}},\ }\href
  {https://doi.org/10.1103/RevModPhys.93.025006} {\bibfield  {journal}
  {\bibinfo  {journal} {Rev. Mod. Phys.}\ }\textbf {\bibinfo {volume} {93}},\
  \bibinfo {pages} {025006} (\bibinfo {year} {2021})}\BibitemShut {NoStop}%
\bibitem [{\citenamefont
  {Mukamel}(1999)}]{mukamelPrinciplesNonlinearOptical1999}%
  \BibitemOpen
  \bibfield  {author} {\bibinfo {author} {\bibfnamefont {S.}~\bibnamefont
  {Mukamel}},\ }\href@noop {} {\emph {\bibinfo {title} {Principles of
  {{Nonlinear Optical Spectroscopy}}}}}\ (\bibinfo  {publisher} {{Oxford
  University Press}},\ \bibinfo {address} {{New York}},\ \bibinfo {year}
  {1999})\BibitemShut {NoStop}%
\bibitem [{\citenamefont {Sodemann}\ and\ \citenamefont
  {Fu}(2015)}]{sodemannQuantumNonlinearHall2015}%
  \BibitemOpen
  \bibfield  {author} {\bibinfo {author} {\bibfnamefont {I.}~\bibnamefont
  {Sodemann}}\ and\ \bibinfo {author} {\bibfnamefont {L.}~\bibnamefont {Fu}},\
  }\href {https://doi.org/10.1103/PhysRevLett.115.216806} {\bibfield  {journal}
  {\bibinfo  {journal} {Phys. Rev. Lett.}\ }\textbf {\bibinfo {volume} {115}},\
  \bibinfo {pages} {216806} (\bibinfo {year} {2015})}\BibitemShut {NoStop}%
\bibitem [{\citenamefont {Ma}\ \emph {et~al.}(2019)\citenamefont {Ma},
  \citenamefont {Xu}, \citenamefont {Shen}, \citenamefont {MacNeill},
  \citenamefont {Fatemi}, \citenamefont {Chang}, \citenamefont {Mier~Valdivia},
  \citenamefont {Wu}, \citenamefont {Du}, \citenamefont {Hsu}, \citenamefont
  {Fang}, \citenamefont {Gibson}, \citenamefont {Watanabe}, \citenamefont
  {Taniguchi}, \citenamefont {Cava}, \citenamefont {Kaxiras}, \citenamefont
  {Lu}, \citenamefont {Lin}, \citenamefont {Fu}, \citenamefont {Gedik},\ and\
  \citenamefont {{Jarillo-Herrero}}}]{maObservationNonlinearHall2019}%
  \BibitemOpen
  \bibfield  {author} {\bibinfo {author} {\bibfnamefont {Q.}~\bibnamefont
  {Ma}}, \bibinfo {author} {\bibfnamefont {S.-Y.}\ \bibnamefont {Xu}}, \bibinfo
  {author} {\bibfnamefont {H.}~\bibnamefont {Shen}}, \bibinfo {author}
  {\bibfnamefont {D.}~\bibnamefont {MacNeill}}, \bibinfo {author}
  {\bibfnamefont {V.}~\bibnamefont {Fatemi}}, \bibinfo {author} {\bibfnamefont
  {T.-R.}\ \bibnamefont {Chang}}, \bibinfo {author} {\bibfnamefont {A.~M.}\
  \bibnamefont {Mier~Valdivia}}, \bibinfo {author} {\bibfnamefont
  {S.}~\bibnamefont {Wu}}, \bibinfo {author} {\bibfnamefont {Z.}~\bibnamefont
  {Du}}, \bibinfo {author} {\bibfnamefont {C.-H.}\ \bibnamefont {Hsu}},
  \bibinfo {author} {\bibfnamefont {S.}~\bibnamefont {Fang}}, \bibinfo {author}
  {\bibfnamefont {Q.~D.}\ \bibnamefont {Gibson}}, \bibinfo {author}
  {\bibfnamefont {K.}~\bibnamefont {Watanabe}}, \bibinfo {author}
  {\bibfnamefont {T.}~\bibnamefont {Taniguchi}}, \bibinfo {author}
  {\bibfnamefont {R.~J.}\ \bibnamefont {Cava}}, \bibinfo {author}
  {\bibfnamefont {E.}~\bibnamefont {Kaxiras}}, \bibinfo {author} {\bibfnamefont
  {H.-Z.}\ \bibnamefont {Lu}}, \bibinfo {author} {\bibfnamefont
  {H.}~\bibnamefont {Lin}}, \bibinfo {author} {\bibfnamefont {L.}~\bibnamefont
  {Fu}}, \bibinfo {author} {\bibfnamefont {N.}~\bibnamefont {Gedik}},\ and\
  \bibinfo {author} {\bibfnamefont {P.}~\bibnamefont {{Jarillo-Herrero}}},\
  }\href {https://doi.org/10.1038/s41586-018-0807-6} {\bibfield  {journal}
  {\bibinfo  {journal} {Nature}\ }\textbf {\bibinfo {volume} {565}},\ \bibinfo
  {pages} {337} (\bibinfo {year} {2019})}\BibitemShut {NoStop}%
\bibitem [{\citenamefont {Lai}\ \emph {et~al.}(2021)\citenamefont {Lai},
  \citenamefont {Liu}, \citenamefont {Zhang}, \citenamefont {Zhao},
  \citenamefont {Feng}, \citenamefont {Wang}, \citenamefont {Tang},
  \citenamefont {Liu}, \citenamefont {Novoselov}, \citenamefont {Yang},\ and\
  \citenamefont {Gao}}]{laiThirdorderNonlinearHall2021}%
  \BibitemOpen
  \bibfield  {author} {\bibinfo {author} {\bibfnamefont {S.}~\bibnamefont
  {Lai}}, \bibinfo {author} {\bibfnamefont {H.}~\bibnamefont {Liu}}, \bibinfo
  {author} {\bibfnamefont {Z.}~\bibnamefont {Zhang}}, \bibinfo {author}
  {\bibfnamefont {J.}~\bibnamefont {Zhao}}, \bibinfo {author} {\bibfnamefont
  {X.}~\bibnamefont {Feng}}, \bibinfo {author} {\bibfnamefont {N.}~\bibnamefont
  {Wang}}, \bibinfo {author} {\bibfnamefont {C.}~\bibnamefont {Tang}}, \bibinfo
  {author} {\bibfnamefont {Y.}~\bibnamefont {Liu}}, \bibinfo {author}
  {\bibfnamefont {K.~S.}\ \bibnamefont {Novoselov}}, \bibinfo {author}
  {\bibfnamefont {S.~A.}\ \bibnamefont {Yang}},\ and\ \bibinfo {author}
  {\bibfnamefont {W.-b.}\ \bibnamefont {Gao}},\ }\href
  {https://doi.org/10.1038/s41565-021-00917-0} {\bibfield  {journal} {\bibinfo
  {journal} {Nat. Nanotechnol.}\ }\textbf {\bibinfo {volume} {16}},\ \bibinfo
  {pages} {869} (\bibinfo {year} {2021})}\BibitemShut {NoStop}%
\bibitem [{\citenamefont {Ahn}\ \emph {et~al.}(2022)\citenamefont {Ahn},
  \citenamefont {Guo}, \citenamefont {Nagaosa},\ and\ \citenamefont
  {Vishwanath}}]{ahnRiemannianGeometryResonant2022}%
  \BibitemOpen
  \bibfield  {author} {\bibinfo {author} {\bibfnamefont {J.}~\bibnamefont
  {Ahn}}, \bibinfo {author} {\bibfnamefont {G.-Y.}\ \bibnamefont {Guo}},
  \bibinfo {author} {\bibfnamefont {N.}~\bibnamefont {Nagaosa}},\ and\ \bibinfo
  {author} {\bibfnamefont {A.}~\bibnamefont {Vishwanath}},\ }\href
  {https://doi.org/10.1038/s41567-021-01465-z} {\bibfield  {journal} {\bibinfo
  {journal} {Nat. Phys.}\ }\textbf {\bibinfo {volume} {18}},\ \bibinfo {pages}
  {290} (\bibinfo {year} {2022})}\BibitemShut {NoStop}%
\bibitem [{\citenamefont {Fiebig}\ \emph {et~al.}(2005)\citenamefont {Fiebig},
  \citenamefont {Pavlov},\ and\ \citenamefont
  {Pisarev}}]{fiebigSecondharmonicGenerationTool2005}%
  \BibitemOpen
  \bibfield  {author} {\bibinfo {author} {\bibfnamefont {M.}~\bibnamefont
  {Fiebig}}, \bibinfo {author} {\bibfnamefont {V.~V.}\ \bibnamefont {Pavlov}},\
  and\ \bibinfo {author} {\bibfnamefont {R.~V.}\ \bibnamefont {Pisarev}},\
  }\href {https://doi.org/10.1364/JOSAB.22.000096} {\bibfield  {journal}
  {\bibinfo  {journal} {J. Opt. Soc. Am. B, JOSAB}\ }\textbf {\bibinfo {volume}
  {22}},\ \bibinfo {pages} {96} (\bibinfo {year} {2005})}\BibitemShut {NoStop}%
\bibitem [{\citenamefont {Zhao}\ \emph {et~al.}(2018)\citenamefont {Zhao},
  \citenamefont {Torchinsky}, \citenamefont {Harter}, \citenamefont {{de la
  Torre}},\ and\ \citenamefont {Hsieh}}]{zhaoSecondHarmonicGeneration2018}%
  \BibitemOpen
  \bibfield  {author} {\bibinfo {author} {\bibfnamefont {L.}~\bibnamefont
  {Zhao}}, \bibinfo {author} {\bibfnamefont {D.}~\bibnamefont {Torchinsky}},
  \bibinfo {author} {\bibfnamefont {J.}~\bibnamefont {Harter}}, \bibinfo
  {author} {\bibfnamefont {A.}~\bibnamefont {{de la Torre}}},\ and\ \bibinfo
  {author} {\bibfnamefont {D.}~\bibnamefont {Hsieh}},\ }in\ \href
  {https://doi.org/10.1016/B978-0-12-803581-8.09533-3} {\emph {\bibinfo
  {booktitle} {Encyclopedia of {{Modern Optics}} ({{Second Edition}})}}},\
  \bibinfo {editor} {edited by\ \bibinfo {editor} {\bibfnamefont {B.~D.}\
  \bibnamefont {Guenther}}\ and\ \bibinfo {editor} {\bibfnamefont {D.~G.}\
  \bibnamefont {Steel}}}\ (\bibinfo  {publisher} {{Elsevier}},\ \bibinfo
  {address} {{Oxford}},\ \bibinfo {year} {2018})\ pp.\ \bibinfo {pages}
  {207--226}\BibitemShut {NoStop}%
\bibitem [{\citenamefont {Sirica}\ \emph {et~al.}(2022)\citenamefont {Sirica},
  \citenamefont {Orth}, \citenamefont {Scheurer}, \citenamefont {Dai},
  \citenamefont {Lee}, \citenamefont {Padmanabhan}, \citenamefont {Mix},
  \citenamefont {Teitelbaum}, \citenamefont {Trigo}, \citenamefont {Zhao},
  \citenamefont {Chen}, \citenamefont {Xu}, \citenamefont {Yang}, \citenamefont
  {Shen}, \citenamefont {Hu}, \citenamefont {Lee}, \citenamefont {Lin},
  \citenamefont {Cochran}, \citenamefont {Trugman}, \citenamefont {Zhu},
  \citenamefont {Hasan}, \citenamefont {Ni}, \citenamefont {Qiu}, \citenamefont
  {Taylor}, \citenamefont {Yarotski},\ and\ \citenamefont
  {Prasankumar}}]{siricaPhotocurrentdrivenTransientSymmetry2022}%
  \BibitemOpen
  \bibfield  {author} {\bibinfo {author} {\bibfnamefont {N.}~\bibnamefont
  {Sirica}}, \bibinfo {author} {\bibfnamefont {P.~P.}\ \bibnamefont {Orth}},
  \bibinfo {author} {\bibfnamefont {M.~S.}\ \bibnamefont {Scheurer}}, \bibinfo
  {author} {\bibfnamefont {Y.~M.}\ \bibnamefont {Dai}}, \bibinfo {author}
  {\bibfnamefont {M.-C.}\ \bibnamefont {Lee}}, \bibinfo {author} {\bibfnamefont
  {P.}~\bibnamefont {Padmanabhan}}, \bibinfo {author} {\bibfnamefont {L.~T.}\
  \bibnamefont {Mix}}, \bibinfo {author} {\bibfnamefont {S.~W.}\ \bibnamefont
  {Teitelbaum}}, \bibinfo {author} {\bibfnamefont {M.}~\bibnamefont {Trigo}},
  \bibinfo {author} {\bibfnamefont {L.~X.}\ \bibnamefont {Zhao}}, \bibinfo
  {author} {\bibfnamefont {G.~F.}\ \bibnamefont {Chen}}, \bibinfo {author}
  {\bibfnamefont {B.}~\bibnamefont {Xu}}, \bibinfo {author} {\bibfnamefont
  {R.}~\bibnamefont {Yang}}, \bibinfo {author} {\bibfnamefont {B.}~\bibnamefont
  {Shen}}, \bibinfo {author} {\bibfnamefont {C.}~\bibnamefont {Hu}}, \bibinfo
  {author} {\bibfnamefont {C.-C.}\ \bibnamefont {Lee}}, \bibinfo {author}
  {\bibfnamefont {H.}~\bibnamefont {Lin}}, \bibinfo {author} {\bibfnamefont
  {T.~A.}\ \bibnamefont {Cochran}}, \bibinfo {author} {\bibfnamefont {S.~A.}\
  \bibnamefont {Trugman}}, \bibinfo {author} {\bibfnamefont {J.-X.}\
  \bibnamefont {Zhu}}, \bibinfo {author} {\bibfnamefont {M.~Z.}\ \bibnamefont
  {Hasan}}, \bibinfo {author} {\bibfnamefont {N.}~\bibnamefont {Ni}}, \bibinfo
  {author} {\bibfnamefont {X.~G.}\ \bibnamefont {Qiu}}, \bibinfo {author}
  {\bibfnamefont {A.~J.}\ \bibnamefont {Taylor}}, \bibinfo {author}
  {\bibfnamefont {D.~A.}\ \bibnamefont {Yarotski}},\ and\ \bibinfo {author}
  {\bibfnamefont {R.~P.}\ \bibnamefont {Prasankumar}},\ }\href
  {https://doi.org/10.1038/s41563-021-01126-9} {\bibfield  {journal} {\bibinfo
  {journal} {Nat. Mater.}\ }\textbf {\bibinfo {volume} {21}},\ \bibinfo {pages}
  {62} (\bibinfo {year} {2022})}\BibitemShut {NoStop}%
\bibitem [{\citenamefont {Hamm}\ and\ \citenamefont
  {Zanni}(2011)}]{hammConceptsMethods2D2011}%
  \BibitemOpen
  \bibfield  {author} {\bibinfo {author} {\bibfnamefont {P.}~\bibnamefont
  {Hamm}}\ and\ \bibinfo {author} {\bibfnamefont {M.}~\bibnamefont {Zanni}},\
  }\href@noop {} {\emph {\bibinfo {title} {Concepts and {{Methods}} of {{2D
  Infrared Spectroscopy}}}}},\ \bibinfo {edition} {illustrated edition}\ ed.\
  (\bibinfo  {publisher} {{Cambridge University Press}},\ \bibinfo {address}
  {{Cambridge ; New York}},\ \bibinfo {year} {2011})\BibitemShut {NoStop}%
\bibitem [{\citenamefont {Lu}\ \emph {et~al.}(2018)\citenamefont {Lu},
  \citenamefont {Li}, \citenamefont {Zhang}, \citenamefont {Hwang},
  \citenamefont {{Ofori-Okai}},\ and\ \citenamefont
  {Nelson}}]{luTwoDimensionalSpectroscopyTerahertz2018}%
  \BibitemOpen
  \bibfield  {author} {\bibinfo {author} {\bibfnamefont {J.}~\bibnamefont
  {Lu}}, \bibinfo {author} {\bibfnamefont {X.}~\bibnamefont {Li}}, \bibinfo
  {author} {\bibfnamefont {Y.}~\bibnamefont {Zhang}}, \bibinfo {author}
  {\bibfnamefont {H.~Y.}\ \bibnamefont {Hwang}}, \bibinfo {author}
  {\bibfnamefont {B.~K.}\ \bibnamefont {{Ofori-Okai}}},\ and\ \bibinfo {author}
  {\bibfnamefont {K.~A.}\ \bibnamefont {Nelson}},\ }\href
  {https://doi.org/10.1007/s41061-018-0185-4} {\bibfield  {journal} {\bibinfo
  {journal} {Top Curr Chem (Z)}\ }\textbf {\bibinfo {volume} {376}},\ \bibinfo
  {pages} {6} (\bibinfo {year} {2018})}\BibitemShut {NoStop}%
\bibitem [{\citenamefont {Kuehn}\ \emph {et~al.}(2011)\citenamefont {Kuehn},
  \citenamefont {Reimann}, \citenamefont {Woerner}, \citenamefont {Elsaesser},\
  and\ \citenamefont {Hey}}]{kuehnTwoDimensionalTerahertzCorrelation2011}%
  \BibitemOpen
  \bibfield  {author} {\bibinfo {author} {\bibfnamefont {W.}~\bibnamefont
  {Kuehn}}, \bibinfo {author} {\bibfnamefont {K.}~\bibnamefont {Reimann}},
  \bibinfo {author} {\bibfnamefont {M.}~\bibnamefont {Woerner}}, \bibinfo
  {author} {\bibfnamefont {T.}~\bibnamefont {Elsaesser}},\ and\ \bibinfo
  {author} {\bibfnamefont {R.}~\bibnamefont {Hey}},\ }\href
  {https://doi.org/10.1021/jp1099046} {\bibfield  {journal} {\bibinfo
  {journal} {J. Phys. Chem. B}\ }\textbf {\bibinfo {volume} {115}},\ \bibinfo
  {pages} {5448} (\bibinfo {year} {2011})}\BibitemShut {NoStop}%
\bibitem [{\citenamefont {Woerner}\ \emph {et~al.}(2013)\citenamefont
  {Woerner}, \citenamefont {Kuehn}, \citenamefont {Bowlan}, \citenamefont
  {Reimann},\ and\ \citenamefont
  {Elsaesser}}]{woernerUltrafastTwodimensionalTerahertz2013}%
  \BibitemOpen
  \bibfield  {author} {\bibinfo {author} {\bibfnamefont {M.}~\bibnamefont
  {Woerner}}, \bibinfo {author} {\bibfnamefont {W.}~\bibnamefont {Kuehn}},
  \bibinfo {author} {\bibfnamefont {P.}~\bibnamefont {Bowlan}}, \bibinfo
  {author} {\bibfnamefont {K.}~\bibnamefont {Reimann}},\ and\ \bibinfo {author}
  {\bibfnamefont {T.}~\bibnamefont {Elsaesser}},\ }\href
  {https://doi.org/10.1088/1367-2630/15/2/025039} {\bibfield  {journal}
  {\bibinfo  {journal} {New J. Phys.}\ }\textbf {\bibinfo {volume} {15}},\
  \bibinfo {pages} {025039} (\bibinfo {year} {2013})}\BibitemShut {NoStop}%
\bibitem [{\citenamefont {Bowlan}\ \emph {et~al.}(2014)\citenamefont {Bowlan},
  \citenamefont {{Martinez-Moreno}}, \citenamefont {Reimann}, \citenamefont
  {Elsaesser},\ and\ \citenamefont
  {Woerner}}]{bowlanUltrafastTerahertzResponse2014}%
  \BibitemOpen
  \bibfield  {author} {\bibinfo {author} {\bibfnamefont {P.}~\bibnamefont
  {Bowlan}}, \bibinfo {author} {\bibfnamefont {E.}~\bibnamefont
  {{Martinez-Moreno}}}, \bibinfo {author} {\bibfnamefont {K.}~\bibnamefont
  {Reimann}}, \bibinfo {author} {\bibfnamefont {T.}~\bibnamefont {Elsaesser}},\
  and\ \bibinfo {author} {\bibfnamefont {M.}~\bibnamefont {Woerner}},\ }\href
  {https://doi.org/10.1103/PhysRevB.89.041408} {\bibfield  {journal} {\bibinfo
  {journal} {Phys. Rev. B}\ }\textbf {\bibinfo {volume} {89}},\ \bibinfo
  {pages} {041408} (\bibinfo {year} {2014})}\BibitemShut {NoStop}%
\bibitem [{\citenamefont {Lu}\ \emph {et~al.}(2017)\citenamefont {Lu},
  \citenamefont {Li}, \citenamefont {Hwang}, \citenamefont {{Ofori-Okai}},
  \citenamefont {Kurihara}, \citenamefont {Suemoto},\ and\ \citenamefont
  {Nelson}}]{luCoherentTwoDimensionalTerahertz2017}%
  \BibitemOpen
  \bibfield  {author} {\bibinfo {author} {\bibfnamefont {J.}~\bibnamefont
  {Lu}}, \bibinfo {author} {\bibfnamefont {X.}~\bibnamefont {Li}}, \bibinfo
  {author} {\bibfnamefont {H.~Y.}\ \bibnamefont {Hwang}}, \bibinfo {author}
  {\bibfnamefont {B.~K.}\ \bibnamefont {{Ofori-Okai}}}, \bibinfo {author}
  {\bibfnamefont {T.}~\bibnamefont {Kurihara}}, \bibinfo {author}
  {\bibfnamefont {T.}~\bibnamefont {Suemoto}},\ and\ \bibinfo {author}
  {\bibfnamefont {K.~A.}\ \bibnamefont {Nelson}},\ }\href
  {https://doi.org/10.1103/PhysRevLett.118.207204} {\bibfield  {journal}
  {\bibinfo  {journal} {Phys. Rev. Lett.}\ }\textbf {\bibinfo {volume} {118}},\
  \bibinfo {pages} {207204} (\bibinfo {year} {2017})}\BibitemShut {NoStop}%
\bibitem [{\citenamefont {Johnson}\ \emph {et~al.}(2019)\citenamefont
  {Johnson}, \citenamefont {Knighton},\ and\ \citenamefont
  {Johnson}}]{johnsonDistinguishingNonlinearTerahertz2019}%
  \BibitemOpen
  \bibfield  {author} {\bibinfo {author} {\bibfnamefont {C.~L.}\ \bibnamefont
  {Johnson}}, \bibinfo {author} {\bibfnamefont {B.~E.}\ \bibnamefont
  {Knighton}},\ and\ \bibinfo {author} {\bibfnamefont {J.~A.}\ \bibnamefont
  {Johnson}},\ }\href {https://doi.org/10.1103/PhysRevLett.122.073901}
  {\bibfield  {journal} {\bibinfo  {journal} {Phys. Rev. Lett.}\ }\textbf
  {\bibinfo {volume} {122}},\ \bibinfo {pages} {073901} (\bibinfo {year}
  {2019})}\BibitemShut {NoStop}%
\bibitem [{\citenamefont {Mahmood}\ \emph {et~al.}(2021)\citenamefont
  {Mahmood}, \citenamefont {Chaudhuri}, \citenamefont {Gopalakrishnan},
  \citenamefont {Nandkishore},\ and\ \citenamefont
  {Armitage}}]{mahmoodObservationMarginalFermi2021}%
  \BibitemOpen
  \bibfield  {author} {\bibinfo {author} {\bibfnamefont {F.}~\bibnamefont
  {Mahmood}}, \bibinfo {author} {\bibfnamefont {D.}~\bibnamefont {Chaudhuri}},
  \bibinfo {author} {\bibfnamefont {S.}~\bibnamefont {Gopalakrishnan}},
  \bibinfo {author} {\bibfnamefont {R.}~\bibnamefont {Nandkishore}},\ and\
  \bibinfo {author} {\bibfnamefont {N.~P.}\ \bibnamefont {Armitage}},\ }\href
  {https://doi.org/10.1038/s41567-020-01149-0} {\bibfield  {journal} {\bibinfo
  {journal} {Nat. Phys.}\ }\textbf {\bibinfo {volume} {17}},\ \bibinfo {pages}
  {627} (\bibinfo {year} {2021})}\BibitemShut {NoStop}%
\bibitem [{\citenamefont {Lin}\ \emph {et~al.}(2022)\citenamefont {Lin},
  \citenamefont {Mead},\ and\ \citenamefont
  {Blake}}]{linMappingMathrmLiNbO32022}%
  \BibitemOpen
  \bibfield  {author} {\bibinfo {author} {\bibfnamefont {H.-W.}\ \bibnamefont
  {Lin}}, \bibinfo {author} {\bibfnamefont {G.}~\bibnamefont {Mead}},\ and\
  \bibinfo {author} {\bibfnamefont {G.~A.}\ \bibnamefont {Blake}},\ }\href
  {https://doi.org/10.1103/PhysRevLett.129.207401} {\bibfield  {journal}
  {\bibinfo  {journal} {Phys. Rev. Lett.}\ }\textbf {\bibinfo {volume} {129}},\
  \bibinfo {pages} {207401} (\bibinfo {year} {2022})}\BibitemShut {NoStop}%
\bibitem [{\citenamefont {Luo}\ \emph {et~al.}(2022)\citenamefont {Luo},
  \citenamefont {Mootz}, \citenamefont {Kang}, \citenamefont {Huang},
  \citenamefont {Eom}, \citenamefont {Lee}, \citenamefont {Vaswani},
  \citenamefont {Collantes}, \citenamefont {Hellstrom}, \citenamefont
  {Perakis}, \citenamefont {Eom},\ and\ \citenamefont
  {Wang}}]{luoQuantumCoherenceTomography2022}%
  \BibitemOpen
  \bibfield  {author} {\bibinfo {author} {\bibfnamefont {L.}~\bibnamefont
  {Luo}}, \bibinfo {author} {\bibfnamefont {M.}~\bibnamefont {Mootz}}, \bibinfo
  {author} {\bibfnamefont {J.~H.}\ \bibnamefont {Kang}}, \bibinfo {author}
  {\bibfnamefont {C.}~\bibnamefont {Huang}}, \bibinfo {author} {\bibfnamefont
  {K.}~\bibnamefont {Eom}}, \bibinfo {author} {\bibfnamefont {J.~W.}\
  \bibnamefont {Lee}}, \bibinfo {author} {\bibfnamefont {C.}~\bibnamefont
  {Vaswani}}, \bibinfo {author} {\bibfnamefont {Y.~G.}\ \bibnamefont
  {Collantes}}, \bibinfo {author} {\bibfnamefont {E.~E.}\ \bibnamefont
  {Hellstrom}}, \bibinfo {author} {\bibfnamefont {I.~E.}\ \bibnamefont
  {Perakis}}, \bibinfo {author} {\bibfnamefont {C.~B.}\ \bibnamefont {Eom}},\
  and\ \bibinfo {author} {\bibfnamefont {J.}~\bibnamefont {Wang}},\ }\href
  {https://doi.org/10.1038/s41567-022-01827-1} {\bibfield  {journal} {\bibinfo
  {journal} {Nat. Phys.}\ ,\ \bibinfo {pages} {1}} (\bibinfo {year}
  {2022})}\BibitemShut {NoStop}%
\bibitem [{\citenamefont {Wan}\ and\ \citenamefont
  {Armitage}(2019)}]{wanResolvingContinuaFractional2019}%
  \BibitemOpen
  \bibfield  {author} {\bibinfo {author} {\bibfnamefont {Y.}~\bibnamefont
  {Wan}}\ and\ \bibinfo {author} {\bibfnamefont {N.~P.}\ \bibnamefont
  {Armitage}},\ }\href {https://doi.org/10.1103/PhysRevLett.122.257401}
  {\bibfield  {journal} {\bibinfo  {journal} {Phys. Rev. Lett.}\ }\textbf
  {\bibinfo {volume} {122}},\ \bibinfo {pages} {257401} (\bibinfo {year}
  {2019})}\BibitemShut {NoStop}%
\bibitem [{\citenamefont {Nandkishore}\ \emph {et~al.}(2021)\citenamefont
  {Nandkishore}, \citenamefont {Choi},\ and\ \citenamefont
  {Kim}}]{nandkishoreSpectroscopicFingerprintsGapped2021}%
  \BibitemOpen
  \bibfield  {author} {\bibinfo {author} {\bibfnamefont {R.~M.}\ \bibnamefont
  {Nandkishore}}, \bibinfo {author} {\bibfnamefont {W.}~\bibnamefont {Choi}},\
  and\ \bibinfo {author} {\bibfnamefont {Y.~B.}\ \bibnamefont {Kim}},\ }\href
  {https://doi.org/10.1103/PhysRevResearch.3.013254} {\bibfield  {journal}
  {\bibinfo  {journal} {Phys. Rev. Research}\ }\textbf {\bibinfo {volume}
  {3}},\ \bibinfo {pages} {013254} (\bibinfo {year} {2021})}\BibitemShut
  {NoStop}%
\bibitem [{\citenamefont {Parameswaran}\ and\ \citenamefont
  {Gopalakrishnan}(2020)}]{parameswaranAsymptoticallyExactTheory2020}%
  \BibitemOpen
  \bibfield  {author} {\bibinfo {author} {\bibfnamefont {S.~A.}\ \bibnamefont
  {Parameswaran}}\ and\ \bibinfo {author} {\bibfnamefont {S.}~\bibnamefont
  {Gopalakrishnan}},\ }\href {https://doi.org/10.1103/PhysRevLett.125.237601}
  {\bibfield  {journal} {\bibinfo  {journal} {Phys. Rev. Lett.}\ }\textbf
  {\bibinfo {volume} {125}},\ \bibinfo {pages} {237601} (\bibinfo {year}
  {2020})}\BibitemShut {NoStop}%
\bibitem [{\citenamefont {Choi}\ \emph {et~al.}(2020)\citenamefont {Choi},
  \citenamefont {Lee},\ and\ \citenamefont
  {Kim}}]{choiTheoryTwoDimensionalNonlinear2020}%
  \BibitemOpen
  \bibfield  {author} {\bibinfo {author} {\bibfnamefont {W.}~\bibnamefont
  {Choi}}, \bibinfo {author} {\bibfnamefont {K.~H.}\ \bibnamefont {Lee}},\ and\
  \bibinfo {author} {\bibfnamefont {Y.~B.}\ \bibnamefont {Kim}},\ }\href
  {https://doi.org/10.1103/PhysRevLett.124.117205} {\bibfield  {journal}
  {\bibinfo  {journal} {Phys. Rev. Lett.}\ }\textbf {\bibinfo {volume} {124}},\
  \bibinfo {pages} {117205} (\bibinfo {year} {2020})}\BibitemShut {NoStop}%
\bibitem [{\citenamefont {Li}\ \emph {et~al.}(2021)\citenamefont {Li},
  \citenamefont {Oshikawa},\ and\ \citenamefont
  {Wan}}]{liPhotonEchoLensing2021}%
  \BibitemOpen
  \bibfield  {author} {\bibinfo {author} {\bibfnamefont {Z.-L.}\ \bibnamefont
  {Li}}, \bibinfo {author} {\bibfnamefont {M.}~\bibnamefont {Oshikawa}},\ and\
  \bibinfo {author} {\bibfnamefont {Y.}~\bibnamefont {Wan}},\ }\href
  {https://doi.org/10.1103/PhysRevX.11.031035} {\bibfield  {journal} {\bibinfo
  {journal} {Phys. Rev. X}\ }\textbf {\bibinfo {volume} {11}},\ \bibinfo
  {pages} {031035} (\bibinfo {year} {2021})}\BibitemShut {NoStop}%
\bibitem [{\citenamefont {Savary}\ and\ \citenamefont
  {Balents}(2017)}]{savaryQuantumSpinLiquids2017}%
  \BibitemOpen
  \bibfield  {author} {\bibinfo {author} {\bibfnamefont {L.}~\bibnamefont
  {Savary}}\ and\ \bibinfo {author} {\bibfnamefont {L.}~\bibnamefont
  {Balents}},\ }\href {https://doi.org/10.1088/0034-4885/80/1/016502}
  {\bibfield  {journal} {\bibinfo  {journal} {Rep. Prog. Phys.}\ }\textbf
  {\bibinfo {volume} {80}},\ \bibinfo {pages} {016502} (\bibinfo {year}
  {2017})}\BibitemShut {NoStop}%
\bibitem [{\citenamefont {Trebst}\ and\ \citenamefont
  {Hickey}(2022)}]{trebstKitaevMaterials2022}%
  \BibitemOpen
  \bibfield  {author} {\bibinfo {author} {\bibfnamefont {S.}~\bibnamefont
  {Trebst}}\ and\ \bibinfo {author} {\bibfnamefont {C.}~\bibnamefont
  {Hickey}},\ }\href {https://doi.org/10.1016/j.physrep.2021.11.003} {\bibfield
   {journal} {\bibinfo  {journal} {Physics Reports}\ }\textbf {\bibinfo
  {volume} {950}},\ \bibinfo {pages} {1} (\bibinfo {year} {2022})}\BibitemShut
  {NoStop}%
\bibitem [{\citenamefont {Chaloupka}\ \emph {et~al.}(2010)\citenamefont
  {Chaloupka}, \citenamefont {Jackeli},\ and\ \citenamefont
  {Khaliullin}}]{chaloupkaKitaevHeisenbergModelHoneycomb2010}%
  \BibitemOpen
  \bibfield  {author} {\bibinfo {author} {\bibfnamefont {J.}~\bibnamefont
  {Chaloupka}}, \bibinfo {author} {\bibfnamefont {G.}~\bibnamefont {Jackeli}},\
  and\ \bibinfo {author} {\bibfnamefont {G.}~\bibnamefont {Khaliullin}},\
  }\href {https://doi.org/10.1103/PhysRevLett.105.027204} {\bibfield  {journal}
  {\bibinfo  {journal} {Phys. Rev. Lett.}\ }\textbf {\bibinfo {volume} {105}},\
  \bibinfo {pages} {027204} (\bibinfo {year} {2010})}\BibitemShut {NoStop}%
\bibitem [{\citenamefont {Kimchi}\ and\ \citenamefont
  {Vishwanath}(2014)}]{kimchiKitaevHeisenbergModelsIridates2014}%
  \BibitemOpen
  \bibfield  {author} {\bibinfo {author} {\bibfnamefont {I.}~\bibnamefont
  {Kimchi}}\ and\ \bibinfo {author} {\bibfnamefont {A.}~\bibnamefont
  {Vishwanath}},\ }\href {https://doi.org/10.1103/PhysRevB.89.014414}
  {\bibfield  {journal} {\bibinfo  {journal} {Phys. Rev. B}\ }\textbf {\bibinfo
  {volume} {89}},\ \bibinfo {pages} {014414} (\bibinfo {year}
  {2014})}\BibitemShut {NoStop}%
\bibitem [{\citenamefont {Liu}\ and\ \citenamefont
  {Khaliullin}(2018)}]{liuPseudospinExchangeInteractions2018}%
  \BibitemOpen
  \bibfield  {author} {\bibinfo {author} {\bibfnamefont {H.}~\bibnamefont
  {Liu}}\ and\ \bibinfo {author} {\bibfnamefont {G.}~\bibnamefont
  {Khaliullin}},\ }\href {https://doi.org/10.1103/PhysRevB.97.014407}
  {\bibfield  {journal} {\bibinfo  {journal} {Phys. Rev. B}\ }\textbf {\bibinfo
  {volume} {97}},\ \bibinfo {pages} {014407} (\bibinfo {year}
  {2018})}\BibitemShut {NoStop}%
\bibitem [{\citenamefont {Sano}\ \emph {et~al.}(2018)\citenamefont {Sano},
  \citenamefont {Kato},\ and\ \citenamefont
  {Motome}}]{sanoKitaevHeisenbergHamiltonianHighspin2018}%
  \BibitemOpen
  \bibfield  {author} {\bibinfo {author} {\bibfnamefont {R.}~\bibnamefont
  {Sano}}, \bibinfo {author} {\bibfnamefont {Y.}~\bibnamefont {Kato}},\ and\
  \bibinfo {author} {\bibfnamefont {Y.}~\bibnamefont {Motome}},\ }\href
  {https://doi.org/10.1103/PhysRevB.97.014408} {\bibfield  {journal} {\bibinfo
  {journal} {Phys. Rev. B}\ }\textbf {\bibinfo {volume} {97}},\ \bibinfo
  {pages} {014408} (\bibinfo {year} {2018})}\BibitemShut {NoStop}%
\bibitem [{\citenamefont {Banerjee}\ \emph {et~al.}(2016)\citenamefont
  {Banerjee}, \citenamefont {Bridges}, \citenamefont {Yan}, \citenamefont
  {Aczel}, \citenamefont {Li}, \citenamefont {Stone}, \citenamefont {Granroth},
  \citenamefont {Lumsden}, \citenamefont {Yiu}, \citenamefont {Knolle},
  \citenamefont {Bhattacharjee}, \citenamefont {Kovrizhin}, \citenamefont
  {Moessner}, \citenamefont {Tennant}, \citenamefont {Mandrus},\ and\
  \citenamefont {Nagler}}]{banerjeeProximateKitaevQuantum2016}%
  \BibitemOpen
  \bibfield  {author} {\bibinfo {author} {\bibfnamefont {A.}~\bibnamefont
  {Banerjee}}, \bibinfo {author} {\bibfnamefont {C.~A.}\ \bibnamefont
  {Bridges}}, \bibinfo {author} {\bibfnamefont {J.-Q.}\ \bibnamefont {Yan}},
  \bibinfo {author} {\bibfnamefont {A.~A.}\ \bibnamefont {Aczel}}, \bibinfo
  {author} {\bibfnamefont {L.}~\bibnamefont {Li}}, \bibinfo {author}
  {\bibfnamefont {M.~B.}\ \bibnamefont {Stone}}, \bibinfo {author}
  {\bibfnamefont {G.~E.}\ \bibnamefont {Granroth}}, \bibinfo {author}
  {\bibfnamefont {M.~D.}\ \bibnamefont {Lumsden}}, \bibinfo {author}
  {\bibfnamefont {Y.}~\bibnamefont {Yiu}}, \bibinfo {author} {\bibfnamefont
  {J.}~\bibnamefont {Knolle}}, \bibinfo {author} {\bibfnamefont
  {S.}~\bibnamefont {Bhattacharjee}}, \bibinfo {author} {\bibfnamefont {D.~L.}\
  \bibnamefont {Kovrizhin}}, \bibinfo {author} {\bibfnamefont {R.}~\bibnamefont
  {Moessner}}, \bibinfo {author} {\bibfnamefont {D.~A.}\ \bibnamefont
  {Tennant}}, \bibinfo {author} {\bibfnamefont {D.~G.}\ \bibnamefont
  {Mandrus}},\ and\ \bibinfo {author} {\bibfnamefont {S.~E.}\ \bibnamefont
  {Nagler}},\ }\href {https://doi.org/10.1038/nmat4604} {\bibfield  {journal}
  {\bibinfo  {journal} {Nature Mater}\ }\textbf {\bibinfo {volume} {15}},\
  \bibinfo {pages} {733} (\bibinfo {year} {2016})}\BibitemShut {NoStop}%
\bibitem [{\citenamefont {Do}\ \emph {et~al.}(2017)\citenamefont {Do},
  \citenamefont {Park}, \citenamefont {Yoshitake}, \citenamefont {Nasu},
  \citenamefont {Motome}, \citenamefont {Kwon}, \citenamefont {Adroja},
  \citenamefont {Voneshen}, \citenamefont {Kim}, \citenamefont {Jang},
  \citenamefont {Park}, \citenamefont {Choi},\ and\ \citenamefont
  {Ji}}]{doMajoranaFermionsKitaev2017}%
  \BibitemOpen
  \bibfield  {author} {\bibinfo {author} {\bibfnamefont {S.-H.}\ \bibnamefont
  {Do}}, \bibinfo {author} {\bibfnamefont {S.-Y.}\ \bibnamefont {Park}},
  \bibinfo {author} {\bibfnamefont {J.}~\bibnamefont {Yoshitake}}, \bibinfo
  {author} {\bibfnamefont {J.}~\bibnamefont {Nasu}}, \bibinfo {author}
  {\bibfnamefont {Y.}~\bibnamefont {Motome}}, \bibinfo {author} {\bibfnamefont
  {Y.~S.}\ \bibnamefont {Kwon}}, \bibinfo {author} {\bibfnamefont {D.~T.}\
  \bibnamefont {Adroja}}, \bibinfo {author} {\bibfnamefont {D.~J.}\
  \bibnamefont {Voneshen}}, \bibinfo {author} {\bibfnamefont {K.}~\bibnamefont
  {Kim}}, \bibinfo {author} {\bibfnamefont {T.-H.}\ \bibnamefont {Jang}},
  \bibinfo {author} {\bibfnamefont {J.-H.}\ \bibnamefont {Park}}, \bibinfo
  {author} {\bibfnamefont {K.-Y.}\ \bibnamefont {Choi}},\ and\ \bibinfo
  {author} {\bibfnamefont {S.}~\bibnamefont {Ji}},\ }\href
  {https://doi.org/10.1038/nphys4264} {\bibfield  {journal} {\bibinfo
  {journal} {Nature Phys}\ }\textbf {\bibinfo {volume} {13}},\ \bibinfo {pages}
  {1079} (\bibinfo {year} {2017})}\BibitemShut {NoStop}%
\bibitem [{\citenamefont {Suzuki}\ \emph {et~al.}(2021)\citenamefont {Suzuki},
  \citenamefont {Liu}, \citenamefont {Bertinshaw}, \citenamefont {Ueda},
  \citenamefont {Kim}, \citenamefont {Laha}, \citenamefont {Weber},
  \citenamefont {Yang}, \citenamefont {Wang}, \citenamefont {Takahashi},
  \citenamefont {F{\"u}rsich}, \citenamefont {Minola}, \citenamefont {Lotsch},
  \citenamefont {Kim}, \citenamefont {Yava{\c s}}, \citenamefont {Daghofer},
  \citenamefont {Chaloupka}, \citenamefont {Khaliullin}, \citenamefont
  {Gretarsson},\ and\ \citenamefont
  {Keimer}}]{suzukiProximateFerromagneticState2021}%
  \BibitemOpen
  \bibfield  {author} {\bibinfo {author} {\bibfnamefont {H.}~\bibnamefont
  {Suzuki}}, \bibinfo {author} {\bibfnamefont {H.}~\bibnamefont {Liu}},
  \bibinfo {author} {\bibfnamefont {J.}~\bibnamefont {Bertinshaw}}, \bibinfo
  {author} {\bibfnamefont {K.}~\bibnamefont {Ueda}}, \bibinfo {author}
  {\bibfnamefont {H.}~\bibnamefont {Kim}}, \bibinfo {author} {\bibfnamefont
  {S.}~\bibnamefont {Laha}}, \bibinfo {author} {\bibfnamefont {D.}~\bibnamefont
  {Weber}}, \bibinfo {author} {\bibfnamefont {Z.}~\bibnamefont {Yang}},
  \bibinfo {author} {\bibfnamefont {L.}~\bibnamefont {Wang}}, \bibinfo {author}
  {\bibfnamefont {H.}~\bibnamefont {Takahashi}}, \bibinfo {author}
  {\bibfnamefont {K.}~\bibnamefont {F{\"u}rsich}}, \bibinfo {author}
  {\bibfnamefont {M.}~\bibnamefont {Minola}}, \bibinfo {author} {\bibfnamefont
  {B.~V.}\ \bibnamefont {Lotsch}}, \bibinfo {author} {\bibfnamefont {B.~J.}\
  \bibnamefont {Kim}}, \bibinfo {author} {\bibfnamefont {H.}~\bibnamefont
  {Yava{\c s}}}, \bibinfo {author} {\bibfnamefont {M.}~\bibnamefont
  {Daghofer}}, \bibinfo {author} {\bibfnamefont {J.}~\bibnamefont {Chaloupka}},
  \bibinfo {author} {\bibfnamefont {G.}~\bibnamefont {Khaliullin}}, \bibinfo
  {author} {\bibfnamefont {H.}~\bibnamefont {Gretarsson}},\ and\ \bibinfo
  {author} {\bibfnamefont {B.}~\bibnamefont {Keimer}},\ }\href
  {https://doi.org/10.1038/s41467-021-24722-4} {\bibfield  {journal} {\bibinfo
  {journal} {Nat Commun}\ }\textbf {\bibinfo {volume} {12}},\ \bibinfo {pages}
  {4512} (\bibinfo {year} {2021})}\BibitemShut {NoStop}%
\bibitem [{\citenamefont {Singh}\ \emph {et~al.}(2012)\citenamefont {Singh},
  \citenamefont {Manni}, \citenamefont {Reuther}, \citenamefont {Berlijn},
  \citenamefont {Thomale}, \citenamefont {Ku}, \citenamefont {Trebst},\ and\
  \citenamefont {Gegenwart}}]{singhRelevanceHeisenbergKitaevModel2012}%
  \BibitemOpen
  \bibfield  {author} {\bibinfo {author} {\bibfnamefont {Y.}~\bibnamefont
  {Singh}}, \bibinfo {author} {\bibfnamefont {S.}~\bibnamefont {Manni}},
  \bibinfo {author} {\bibfnamefont {J.}~\bibnamefont {Reuther}}, \bibinfo
  {author} {\bibfnamefont {T.}~\bibnamefont {Berlijn}}, \bibinfo {author}
  {\bibfnamefont {R.}~\bibnamefont {Thomale}}, \bibinfo {author} {\bibfnamefont
  {W.}~\bibnamefont {Ku}}, \bibinfo {author} {\bibfnamefont {S.}~\bibnamefont
  {Trebst}},\ and\ \bibinfo {author} {\bibfnamefont {P.}~\bibnamefont
  {Gegenwart}},\ }\href {https://doi.org/10.1103/PhysRevLett.108.127203}
  {\bibfield  {journal} {\bibinfo  {journal} {Phys. Rev. Lett.}\ }\textbf
  {\bibinfo {volume} {108}},\ \bibinfo {pages} {127203} (\bibinfo {year}
  {2012})}\BibitemShut {NoStop}%
\bibitem [{\citenamefont {Williams}\ \emph {et~al.}(2016)\citenamefont
  {Williams}, \citenamefont {Johnson}, \citenamefont {Freund}, \citenamefont
  {Choi}, \citenamefont {Jesche}, \citenamefont {Kimchi}, \citenamefont
  {Manni}, \citenamefont {Bombardi}, \citenamefont {Manuel}, \citenamefont
  {Gegenwart},\ and\ \citenamefont
  {Coldea}}]{williamsIncommensurateCounterrotatingMagnetic2016}%
  \BibitemOpen
  \bibfield  {author} {\bibinfo {author} {\bibfnamefont {S.~C.}\ \bibnamefont
  {Williams}}, \bibinfo {author} {\bibfnamefont {R.~D.}\ \bibnamefont
  {Johnson}}, \bibinfo {author} {\bibfnamefont {F.}~\bibnamefont {Freund}},
  \bibinfo {author} {\bibfnamefont {S.}~\bibnamefont {Choi}}, \bibinfo {author}
  {\bibfnamefont {A.}~\bibnamefont {Jesche}}, \bibinfo {author} {\bibfnamefont
  {I.}~\bibnamefont {Kimchi}}, \bibinfo {author} {\bibfnamefont
  {S.}~\bibnamefont {Manni}}, \bibinfo {author} {\bibfnamefont
  {A.}~\bibnamefont {Bombardi}}, \bibinfo {author} {\bibfnamefont
  {P.}~\bibnamefont {Manuel}}, \bibinfo {author} {\bibfnamefont
  {P.}~\bibnamefont {Gegenwart}},\ and\ \bibinfo {author} {\bibfnamefont
  {R.}~\bibnamefont {Coldea}},\ }\href
  {https://doi.org/10.1103/PhysRevB.93.195158} {\bibfield  {journal} {\bibinfo
  {journal} {Phys. Rev. B}\ }\textbf {\bibinfo {volume} {93}},\ \bibinfo
  {pages} {195158} (\bibinfo {year} {2016})}\BibitemShut {NoStop}%
\bibitem [{\citenamefont {Revelli}\ \emph {et~al.}(2020)\citenamefont
  {Revelli}, \citenamefont {Moretti~Sala}, \citenamefont {Monaco},
  \citenamefont {Hickey}, \citenamefont {Becker}, \citenamefont {Freund},
  \citenamefont {Jesche}, \citenamefont {Gegenwart}, \citenamefont {Eschmann},
  \citenamefont {Buessen}, \citenamefont {Trebst}, \citenamefont {{van
  Loosdrecht}}, \citenamefont {{van den Brink}},\ and\ \citenamefont
  {Gr{\"u}ninger}}]{revelliFingerprintsKitaevPhysics2020}%
  \BibitemOpen
  \bibfield  {author} {\bibinfo {author} {\bibfnamefont {A.}~\bibnamefont
  {Revelli}}, \bibinfo {author} {\bibfnamefont {M.}~\bibnamefont
  {Moretti~Sala}}, \bibinfo {author} {\bibfnamefont {G.}~\bibnamefont
  {Monaco}}, \bibinfo {author} {\bibfnamefont {C.}~\bibnamefont {Hickey}},
  \bibinfo {author} {\bibfnamefont {P.}~\bibnamefont {Becker}}, \bibinfo
  {author} {\bibfnamefont {F.}~\bibnamefont {Freund}}, \bibinfo {author}
  {\bibfnamefont {A.}~\bibnamefont {Jesche}}, \bibinfo {author} {\bibfnamefont
  {P.}~\bibnamefont {Gegenwart}}, \bibinfo {author} {\bibfnamefont
  {T.}~\bibnamefont {Eschmann}}, \bibinfo {author} {\bibfnamefont {F.~L.}\
  \bibnamefont {Buessen}}, \bibinfo {author} {\bibfnamefont {S.}~\bibnamefont
  {Trebst}}, \bibinfo {author} {\bibfnamefont {P.~H.~M.}\ \bibnamefont {{van
  Loosdrecht}}}, \bibinfo {author} {\bibfnamefont {J.}~\bibnamefont {{van den
  Brink}}},\ and\ \bibinfo {author} {\bibfnamefont {M.}~\bibnamefont
  {Gr{\"u}ninger}},\ }\href {https://doi.org/10.1103/PhysRevResearch.2.043094}
  {\bibfield  {journal} {\bibinfo  {journal} {Phys. Rev. Research}\ }\textbf
  {\bibinfo {volume} {2}},\ \bibinfo {pages} {043094} (\bibinfo {year}
  {2020})}\BibitemShut {NoStop}%
\bibitem [{\citenamefont {Liu}\ \emph {et~al.}(2020)\citenamefont {Liu},
  \citenamefont {Chaloupka},\ and\ \citenamefont
  {Khaliullin}}]{liuKitaevSpinLiquid2020}%
  \BibitemOpen
  \bibfield  {author} {\bibinfo {author} {\bibfnamefont {H.}~\bibnamefont
  {Liu}}, \bibinfo {author} {\bibfnamefont {J.}~\bibnamefont {Chaloupka}},\
  and\ \bibinfo {author} {\bibfnamefont {G.}~\bibnamefont {Khaliullin}},\
  }\href {https://doi.org/10.1103/PhysRevLett.125.047201} {\bibfield  {journal}
  {\bibinfo  {journal} {Phys. Rev. Lett.}\ }\textbf {\bibinfo {volume} {125}},\
  \bibinfo {pages} {047201} (\bibinfo {year} {2020})}\BibitemShut {NoStop}%
\bibitem [{\citenamefont {Zhang}\ \emph {et~al.}(2023)\citenamefont {Zhang},
  \citenamefont {Xu}, \citenamefont {Halloran}, \citenamefont {Zhong},
  \citenamefont {Broholm}, \citenamefont {Cava}, \citenamefont {Drichko},\ and\
  \citenamefont {Armitage}}]{zhangMagneticContinuumCobaltbased2023}%
  \BibitemOpen
  \bibfield  {author} {\bibinfo {author} {\bibfnamefont {X.}~\bibnamefont
  {Zhang}}, \bibinfo {author} {\bibfnamefont {Y.}~\bibnamefont {Xu}}, \bibinfo
  {author} {\bibfnamefont {T.}~\bibnamefont {Halloran}}, \bibinfo {author}
  {\bibfnamefont {R.}~\bibnamefont {Zhong}}, \bibinfo {author} {\bibfnamefont
  {C.}~\bibnamefont {Broholm}}, \bibinfo {author} {\bibfnamefont {R.~J.}\
  \bibnamefont {Cava}}, \bibinfo {author} {\bibfnamefont {N.}~\bibnamefont
  {Drichko}},\ and\ \bibinfo {author} {\bibfnamefont {N.~P.}\ \bibnamefont
  {Armitage}},\ }\href {https://doi.org/10.1038/s41563-022-01403-1} {\bibfield
  {journal} {\bibinfo  {journal} {Nat. Mater.}\ }\textbf {\bibinfo {volume}
  {22}},\ \bibinfo {pages} {58} (\bibinfo {year} {2023})}\BibitemShut {NoStop}%
\bibitem [{\citenamefont {Halloran}\ \emph {et~al.}(2023)\citenamefont
  {Halloran}, \citenamefont {Desrochers}, \citenamefont {Zhang}, \citenamefont
  {Chen}, \citenamefont {Chern}, \citenamefont {Xu}, \citenamefont {Winn},
  \citenamefont {{Graves-Brook}}, \citenamefont {Stone}, \citenamefont
  {Kolesnikov}, \citenamefont {Qiu}, \citenamefont {Zhong}, \citenamefont
  {Cava}, \citenamefont {Kim},\ and\ \citenamefont
  {Broholm}}]{halloranGeometricalFrustrationKitaev2023}%
  \BibitemOpen
  \bibfield  {author} {\bibinfo {author} {\bibfnamefont {T.}~\bibnamefont
  {Halloran}}, \bibinfo {author} {\bibfnamefont {F.}~\bibnamefont
  {Desrochers}}, \bibinfo {author} {\bibfnamefont {E.~Z.}\ \bibnamefont
  {Zhang}}, \bibinfo {author} {\bibfnamefont {T.}~\bibnamefont {Chen}},
  \bibinfo {author} {\bibfnamefont {L.~E.}\ \bibnamefont {Chern}}, \bibinfo
  {author} {\bibfnamefont {Z.}~\bibnamefont {Xu}}, \bibinfo {author}
  {\bibfnamefont {B.}~\bibnamefont {Winn}}, \bibinfo {author} {\bibfnamefont
  {M.}~\bibnamefont {{Graves-Brook}}}, \bibinfo {author} {\bibfnamefont
  {M.~B.}\ \bibnamefont {Stone}}, \bibinfo {author} {\bibfnamefont {A.~I.}\
  \bibnamefont {Kolesnikov}}, \bibinfo {author} {\bibfnamefont
  {Y.}~\bibnamefont {Qiu}}, \bibinfo {author} {\bibfnamefont {R.}~\bibnamefont
  {Zhong}}, \bibinfo {author} {\bibfnamefont {R.}~\bibnamefont {Cava}},
  \bibinfo {author} {\bibfnamefont {Y.~B.}\ \bibnamefont {Kim}},\ and\ \bibinfo
  {author} {\bibfnamefont {C.}~\bibnamefont {Broholm}},\ }\href
  {https://doi.org/10.1073/pnas.2215509119} {\bibfield  {journal} {\bibinfo
  {journal} {Proceedings of the National Academy of Sciences}\ }\textbf
  {\bibinfo {volume} {120}},\ \bibinfo {pages} {e2215509119} (\bibinfo {year}
  {2023})}\BibitemShut {NoStop}%
\bibitem [{\citenamefont {Tu}\ \emph {et~al.}(2023)\citenamefont {Tu},
  \citenamefont {Dai}, \citenamefont {Zhang}, \citenamefont {Zhao},
  \citenamefont {Jin}, \citenamefont {Gao}, \citenamefont {Chen}, \citenamefont
  {Dai},\ and\ \citenamefont {Li}}]{tuEvidenceGaplessQuantum2023}%
  \BibitemOpen
  \bibfield  {author} {\bibinfo {author} {\bibfnamefont {C.}~\bibnamefont
  {Tu}}, \bibinfo {author} {\bibfnamefont {D.}~\bibnamefont {Dai}}, \bibinfo
  {author} {\bibfnamefont {X.}~\bibnamefont {Zhang}}, \bibinfo {author}
  {\bibfnamefont {C.}~\bibnamefont {Zhao}}, \bibinfo {author} {\bibfnamefont
  {X.}~\bibnamefont {Jin}}, \bibinfo {author} {\bibfnamefont {B.}~\bibnamefont
  {Gao}}, \bibinfo {author} {\bibfnamefont {T.}~\bibnamefont {Chen}}, \bibinfo
  {author} {\bibfnamefont {P.}~\bibnamefont {Dai}},\ and\ \bibinfo {author}
  {\bibfnamefont {S.}~\bibnamefont {Li}},\ }\href
  {https://doi.org/10.48550/arXiv.2212.07322} {\bibinfo {title} {Evidence for
  gapless quantum spin liquid in a honeycomb lattice}} (\bibinfo {year}
  {2023}),\ \Eprint {https://arxiv.org/abs/2212.07322} {arXiv:2212.07322
  [cond-mat]} \BibitemShut {NoStop}%
\bibitem [{\citenamefont {Kitaev}(2006)}]{kitaevAnyonsExactlySolved2006}%
  \BibitemOpen
  \bibfield  {author} {\bibinfo {author} {\bibfnamefont {A.}~\bibnamefont
  {Kitaev}},\ }\href {https://doi.org/10.1016/j.aop.2005.10.005} {\bibfield
  {journal} {\bibinfo  {journal} {Annals of Physics}\ }\textbf {\bibinfo
  {volume} {321}},\ \bibinfo {pages} {2} (\bibinfo {year} {2006})}\BibitemShut
  {NoStop}%
\bibitem [{\citenamefont {Baskaran}\ \emph {et~al.}(2007)\citenamefont
  {Baskaran}, \citenamefont {Mandal},\ and\ \citenamefont
  {Shankar}}]{baskaranExactResultsSpin2007}%
  \BibitemOpen
  \bibfield  {author} {\bibinfo {author} {\bibfnamefont {G.}~\bibnamefont
  {Baskaran}}, \bibinfo {author} {\bibfnamefont {S.}~\bibnamefont {Mandal}},\
  and\ \bibinfo {author} {\bibfnamefont {R.}~\bibnamefont {Shankar}},\ }\href
  {https://doi.org/10.1103/PhysRevLett.98.247201} {\bibfield  {journal}
  {\bibinfo  {journal} {Phys. Rev. Lett.}\ }\textbf {\bibinfo {volume} {98}},\
  \bibinfo {pages} {247201} (\bibinfo {year} {2007})}\BibitemShut {NoStop}%
\bibitem [{\citenamefont {Knolle}\ \emph {et~al.}(2014)\citenamefont {Knolle},
  \citenamefont {Kovrizhin}, \citenamefont {Chalker},\ and\ \citenamefont
  {Moessner}}]{knolleDynamicsTwoDimensionalQuantum2014}%
  \BibitemOpen
  \bibfield  {author} {\bibinfo {author} {\bibfnamefont {J.}~\bibnamefont
  {Knolle}}, \bibinfo {author} {\bibfnamefont {D.~L.}\ \bibnamefont
  {Kovrizhin}}, \bibinfo {author} {\bibfnamefont {J.~T.}\ \bibnamefont
  {Chalker}},\ and\ \bibinfo {author} {\bibfnamefont {R.}~\bibnamefont
  {Moessner}},\ }\href {https://doi.org/10.1103/PhysRevLett.112.207203}
  {\bibfield  {journal} {\bibinfo  {journal} {Phys. Rev. Lett.}\ }\textbf
  {\bibinfo {volume} {112}},\ \bibinfo {pages} {207203} (\bibinfo {year}
  {2014})}\BibitemShut {NoStop}%
\bibitem [{\citenamefont {Qiang}\ \emph {et~al.}(2023)\citenamefont {Qiang},
  \citenamefont {Quito}, \citenamefont {Trevisan},\ and\ \citenamefont
  {Orth}}]{qiangSupplementaryMaterial2023}%
  \BibitemOpen
  \bibfield  {author} {\bibinfo {author} {\bibfnamefont {Y.}~\bibnamefont
  {Qiang}}, \bibinfo {author} {\bibfnamefont {V.~L.}\ \bibnamefont {Quito}},
  \bibinfo {author} {\bibfnamefont {T.~V.}\ \bibnamefont {Trevisan}},\ and\
  \bibinfo {author} {\bibfnamefont {P.~P.}\ \bibnamefont {Orth}},\ }\href@noop
  {} {\bibinfo {title} {Supplementary {{Material}}}} (\bibinfo {year}
  {2023})\BibitemShut {NoStop}%
\bibitem [{\citenamefont {Zschocke}\ and\ \citenamefont
  {Vojta}(2015)}]{zschockePhysicalStatesFinitesize2015}%
  \BibitemOpen
  \bibfield  {author} {\bibinfo {author} {\bibfnamefont {F.}~\bibnamefont
  {Zschocke}}\ and\ \bibinfo {author} {\bibfnamefont {M.}~\bibnamefont
  {Vojta}},\ }\href {https://doi.org/10.1103/PhysRevB.92.014403} {\bibfield
  {journal} {\bibinfo  {journal} {Phys. Rev. B}\ }\textbf {\bibinfo {volume}
  {92}},\ \bibinfo {pages} {014403} (\bibinfo {year} {2015})}\BibitemShut
  {NoStop}%
\bibitem [{\citenamefont {Blaizot}\ and\ \citenamefont
  {Ripka}(1985)}]{blaizotQuantumTheoryFinite1985}%
  \BibitemOpen
  \bibfield  {author} {\bibinfo {author} {\bibfnamefont {J.}~\bibnamefont
  {Blaizot}}\ and\ \bibinfo {author} {\bibfnamefont {G.}~\bibnamefont
  {Ripka}},\ }\href@noop {} {\emph {\bibinfo {title} {Quantum {{Theory}} of
  {{Finite Systems}}}}}\ (\bibinfo  {publisher} {{The MIT Press}},\ \bibinfo
  {address} {{Cambridge, Massachusetts}},\ \bibinfo {year} {1985})\BibitemShut
  {NoStop}%
\bibitem [{\citenamefont {Knolle}(2016)}]{knolleDynamicsQuantumSpin2016}%
  \BibitemOpen
  \bibfield  {author} {\bibinfo {author} {\bibfnamefont {J.}~\bibnamefont
  {Knolle}},\ }\href {https://doi.org/10.1007/978-3-319-23953-8} {\emph
  {\bibinfo {title} {Dynamics of a {{Quantum Spin Liquid}}}}},\ Springer
  {{Theses}}\ (\bibinfo  {publisher} {{Springer International Publishing}},\
  \bibinfo {address} {{Cham}},\ \bibinfo {year} {2016})\BibitemShut {NoStop}%
\end{thebibliography}
\end{document}